\documentclass[letterpaper,11pt]{article}
\usepackage{amsthm}  
\usepackage{amsmath, amssymb}
\usepackage{color}
\usepackage{fullpage}
\usepackage[numbers]{natbib}
\usepackage[noend]{algorithmic}
\usepackage{thmtools, thm-restate}
\usepackage{tikz}
\usepackage{enumerate}
\usepackage{thm-restate}
\usepackage{graphicx}
\usepackage{amsopn}
\usepackage{caption}
\usepackage{hyperref}
\usepackage{subcaption}
\usepackage{zerosum,csquotes}
\usepackage{enumitem,linegoal}
\usepackage[linesnumbered,ruled,vlined]{algorithm2e}
\usepackage{comment}
\usepackage{float}

\usepackage{ctable}					

\usepackage{mathtools}
\usepackage[flushleft]{threeparttable}
\usepackage[normalem]{ulem}

\usepackage{footnote}
\makesavenoteenv{tabular}
\makesavenoteenv{table}

\usepackage[margin=1in]{geometry}

 \newtheorem{theorem}{Theorem}
 \newtheorem{lemma}{Lemma}

\theoremstyle{definition}
 \newtheorem{definition}{Definition}
\newtheorem{remark}{Remark}
\makeatletter
\newif\ifqed
\def\GrabProofArgument[#1]{ #1: \egroup\ignorespaces}
\def\proof{\noindent\textbf\bgroup Proof%
	\@ifnextchar[{\GrabProofArgument}{. \egroup\ignorespaces}\global\qedtrue}

\def\qedhere{\ifmmode\tag*{\qedsign}\else\hspace*{\fill}\qedsign\medskip\fi\global\qedfalse}
\def\qedsign{$\Box$}
\makeatother

\definecolor{mygreen}{RGB}{20,140,80}
\definecolor{mylightgray}{RGB}{230,230,230}

\definecolor{mygreen}{RGB}{20,140,80}
\definecolor{mydarkgray}{gray}{0.15} 
\definecolor{oceanblue}{HTML}{2c55c2}
\hypersetup{
     colorlinks=true,
     citecolor= mygreen,
     linkcolor= black
}

\SetCommentSty{mycommfont}

\newcounter{proccnt}

\DeclareMathOperator*{\argmin}{arg\,min}
\newcommand{\konote}[1]{}

\title{Dynamic Information Design: A Simple Problem on Optimal Sequential Information Disclosure}
\author{Farzaneh Farhadi$^{1,2}$ and Demosthenis Teneketzis$^1$ \\ 
$^1$ University of Michigan, Ann Arbor, USA \\
$^2$ Imperial College London, London, UK \\
Emails: f.farhadi@imperial.ac.uk, teneket@umich.edu
}

\begin{document}
	\newcommand{\ignore}[1]{}
\renewcommand{\theenumi}{(\roman{enumi})}
\renewcommand{\labelenumi}{\theenumi.}
\sloppy

%
%

\date{}

\maketitle


\begin{abstract}
We study a dynamic information design problem in a finite-horizon setting consisting of two strategic and long-term optimizing agents, namely a principal (he) and a detector (she). The principal observes the evolution of a Markov chain that has two states, one ``good'' and one ``bad'' absorbing state, and has to decide how to sequentially disclose information to the detector. The detector's only information consists of the messages she receives from the principal. The detector's objective is to detect as accurately as possible the time of the jump from the good to the bad state. The principal's objective is to delay the detector as much as possible from detective the jump to the bad state. For this setting, we determine the optimal strategies of the principal and the detector. The detector's optimal strategy is described by time-varying thresholds on her posterior belief of the good state. We prove that it is optimal for the principal to give no information to the detector before a time threshold, run a mixed strategy to confuse the detector at the threshold time, and reveal the true state afterwards. We present an algorithm that determines both the optimal time threshold and the optimal mixed strategy that could be employed by the principal. We show, through numerical experiments, that this optimal sequential mechanism significantly outperforms any other information disclosure strategy presented in literature.
\end{abstract}
\section{Introduction}\label{sec:intro}
The decentralization of information is an inevitable facet of managing a large system. In modern technological systems, agents constantly face the challenge of making decisions under incomplete and asymmetric information. This challenge arises in many different applications, including transportation, cyber-security, communication, energy management, smart grids and E-commerce.

There is a large body of literature on the issues related to decision making in informationally decentralized systems when the agents are cooperative and jointly wish to maximize a social welfare function which is usually the sum of their utilities \cite{VECIANA1998,Etzion2005}. The problem is more challenging when the agents are selfish and not concerned about the society as a whole, but only aim at maximizing their own utility. Currently there exist two approaches to the design of efficient multi-agent systems with asymmetric information and selfish/strategic agents. In the following, we briefly describe these approaches.

\emph{1. Mechanism Design.} In this approach, each of the strategic agents (e.g. power generating companies in an energy network) is assumed to possess some private information (e.g. their own valuation/utility functions). There is a coordinator (e.g. an Independent System Operator, ISO) who wishes to optimize a network performance metric (e.g. social welfare) which depends on the agents' private information. To elicit strategic agents' true information, the coordinator (he) needs to provide incentives to them (e.g. a plan for energy production along with monetary taxes or subsidies) so as to align their strategic objectives (e.g. maximization of a strategic agent's utility) with his own objective (e.g. maximization of social welfare). In this situation, the system's information structure (who knows what and when) is fixed, and the coordinator's goal is to design a mechanism that incentivizes the strategic agents to reveal their information truthfully \cite{Rasouli2014}.

\emph{2. Information Design.} In this approach, the coordinator (e.g. a central control center in a metropolitan area) knows perfectly the evolution of the system's state (e.g. the evolution of traffic in the transportation network), but the decision making (e.g. route selection) is done by the strategic agents (e.g. drivers) who have incomplete/imperfect knowledge of the state (e.g. drivers only know the traffic in their immediate vicinity). To incentivize strategic agents to take actions that are desirable for the coordinator, he can provide, sequentially over time, information about the system's state to them (e.g. provide traffic updates to selfish drivers). The goal of information provision/disclosure is the alignment of each agent's objective (e.g. minimization of the average time to reach the destination) with the coordinator's objective (e.g. minimization of the traffic in the roads). In this situation, the game-form/mechanism is fixed but the system's information structure is not fixed. It has to be designed by the coordinator through the sequential disclosure/provision of information to the strategic agents so as to serve his goal \cite{Kamenica2017,Hamid-Allerton}.

This paper addresses an information design problem. Information design problems are also referred to as \emph{Bayesian persuasion} problems because the strategic agents are assumed to understand how information is generated/ manipulated and react to information in a rational (Bayesian) manner. Therefore, information design problems can be seen as persuading Bayesian agents to act in desirable ways. The coordinator who would like to lead other agents to act as he wants is usually referred to as the principal. The system state is the principal's private information which can be used to persuade others to serve his goal.

Information design problems in dynamic environments involving a principal and \emph{long-term-optimizing} strategic agents are challenging to the following reason. The information disclosed by the principal at each time instant impacts the agents' decision making in two ways: 1) it affects the agents' future decisions through altering their knowledge about the principal's private information; 2) it influences their decisions at the past times by changing what they expect to achieve in the future. Currently, very little is known about the design of optimal sequential information disclosure policies in dynamic environments with long-term-optimizing agents. Therefore, we focus on a very simple problem that allows us to highlight how sequential information provision/disclosure strategies can be designed in dynamic environments with long-term optimizing strategic agents.

We consider a version of the quickest detection problem \cite{Shiryaev} with a strategic principal (he) and one strategic agent/detector (she). The detector wants to detect when a two-state discrete-time Markov chain jumps from the ``good'' state to the ``bad'' absorbing state. The detector cannot make direct observations of the Markov chain's state. Instead, she receives, at each time instant, a message from the principal about the Markov chain's state. The principal observes perfectly the evolution of the Markov chain; his objective is to delay, as much as possible, detection of the jump by the detector. At the beginning of the process, the principal commits to a sequential information disclosure strategy/policy which he announces to the detector. The detector’s knowledge of the policy shapes her interpretation of the messages she receives. For each fixed sequential information disclosure policy of the principal, the detector is faced with a standard quickest detection problem with noisy observations. The principal's goal is to determine the sequential information disclosure policy that convinces the detector to wait as long as possible before declaring the jumps. A precise formulation of this problem is presented in Sections \ref{sec:model} and \ref{sec:problem}.

In this paper, we discover an optimal sequential information disclosure strategy for the principal.
We prove that it is optimal for the principal to give no information to the detector before a time threshold, run a mixed strategy to confuse the detector at the threshold time, and reveal the true state afterwards. We present an algorithm that determines both the optimal time threshold and the optimal mixed strategy that could be employed by the principal.

\subsection{Review of Related Works} \label{sec:literature}

Information disclosure mechanisms can be seen as a communication protocol between an information transmitter (he) and one or multiple receivers. A significant part of the existing works in this area have studied the nonstrategic case, where the information transmitter and receivers are cooperative and jointly wish to maximize the global utility \cite{Witsenhausen79,Pravin83, Aditiya2009,Ashutosh1,Demos2006}.

The strategic case of information disclosure mechanisms (Bayesian Persuasion), where the transmitter (principal) and receiver have misaligned objectives, is in tradition of cheap talk \cite{Crawford-Sobel, Battaglini2002, Yuksel2015, Multi-sender2008, Dynamic-cheap-talk2015,Ivanov2016}, and signaling games \cite{Spence,Signaling2017}. In signaling games, the transmitter's utility depends not only on the receiver's actions, but also on his type (private information). However, in information design problems neither the principal's type nor his actions enter his utility directly; they only influence his utility through the effect they have on the receivers' actions. This feature of the model enables us to investigate/analyze the pure effect of principal's private information on the receivers' behaviors.

The main difference between cheap talk and information design is the level of commitment power they give to the principal. In the cheap talk literature, the principal has \emph{no commitment power}; he decides on the information message to be sent after seeing the realization of his private information. The cheap talk model induces a simultaneous game between the principal and the agents. Thus, the main goal of this strand of works is to characterize the (Nash) equilibria of the induced game. Most of the existing work has focused on the static setting \cite{Crawford-Sobel, Battaglini2002, Yuksel2015, Multi-sender2008}. The work of \cite{Crawford-Sobel} shows that when the state of the system (principal's private information) is one-dimensional and both the principal’s and the agent’s utilities are quadratic, the principal’s equilibrium strategy employs quantization. More general models of cheap talk, such as multidimensional sources \cite{Battaglini2002}, noisy communication \cite{Yuksel2015} and multiple principals with misaligned objectives \cite{Multi-sender2008}, have been studied in the literature for static settings. There are a few works that study the dynamic version of the cheap talk communication \cite{Dynamic-cheap-talk2015,Ivanov2016}. These works show that allowing for dynamic information transmission improves the informativeness of communication.

In information design problems the transmitter is endowed with \emph{full commitment power}. In this model, the transmitter is allowed to send any distribution of messages as a function of the system state, but he should choose and announce his information disclosure policy before observing the system state and then stay committed to it forever. By committing to an information disclosure policy at the very beginning of the system's operation, the transmitter can persuade all other agents to employ strategies that achieve his objective. The fact that a player in a game can improve his outcome through commitment was first established in \cite{Stackelberg1934,Schelling1956,schelling1980} (see also \cite{KALAI2010}). The full commitment assumption holds in many economic (\cite{Goldstein2018,Inostroza2018,price-discrimination2015}) and engineering (\cite{Kamenica2017,Hamid-Allerton}) multi-agent systems.


The literature on information design is generally divided into two main categories: \emph{static} and \emph{dynamic} information design problems. 

\textbf{Static information design problems}. The static version of the problem, where the state of the system is fixed and no time is involved, has been studied extensively in the literature. The authors of \cite{Cedric2017} consider a problem of static information disclosure where the state of the system is Gaussian and the utilities are quadratic, and show that there exists a linear policy for the principal that leads to a Stackelberg equilibrium. In \cite{Bayesian-persuasion, Unified-approach} the authors propose a concavification method for deriving an optimal information provision mechanism in static settings. The concavification approach was first developed by a group of U.S. economists and game theorists, led by R. Aumann and M. Maschler, with the context of the negotiations for the Strategic Arms Limitation Treaty (SALT) between the U.S. and U.S.S.R. (see \cite{Aumann1995}). The approach was developed for problems that can be modeled as repeated games of incomplete information. For over half a century, this idea played an important role in the analysis of repeated games (see \cite{ZAMIR1992,FORGES1992} and references therein) but has not been applied to information design problems until 2011 \cite{Bayesian-persuasion}. 

Following \cite{Bayesian-persuasion}, information design problem has been studied for more general settings, such as costly communication \cite{costly2015}, multi-dimensional state \cite{Tamura2012}, multiple principals \cite{multiple-senders2017,multiple-senders2018}, multiple receivers \cite{Bergemann-multiagent,Bergemann2016}, and receivers with different prior beliefs \cite{heterogeneous-priors2016}. A case of information design problem with transfers where the principal can provide both information and monetary incentives to the receiver is also studied in \cite{with-transfer2017}. There is a group of works in static information disclosure which are more applied and aim to understand or improve real-world institutions via information design. Research in this strand includes applications to grading in schools \cite{Grading2015}, research procurement \cite{research-procurement2019}, medical testing \cite{medical-testing2018}, price discrimination \cite{price-discrimination2015}, insurance \cite{insurance2018}, and routing software \cite{Routing2014,Kamenica2017,Hamid-Allerton,Bayesian-Persuasion-routing-game,Saurabh2019}. A through discussion of the literature on information design up until 2018 appears in \cite{Kamenica-survey}.

\textbf{Dynamic information design problems}. The dynamic version of the information design problem, where the informed-player/principal can disclose his superior information sequentially over time, has recently been attracting rapidly growing interest. In dynamic environments, agents' decisions at each instant of time affect their opponents' decisions, not only at present but also in the future. The problem becomes more tangled if the information transmitter has commitment power. In this case, the information disclosure policy that the principal commits to for the future, has a direct effect on the receivers' estimation of what they can gain in the future, and hence on their current decisions. The interdependency among the agents' decision making processes over time makes the dynamic information design problems very complex and challenging.

Most of the available works avoid this challenge by assuming that the agents are myopic, that is they only look at each instant at the immediate consequence of their actions, ignoring the subsequent (future) effects. In \cite{Farokhi2017} and \cite{Basar2016-tracking} both the information transmitter and receivers are assumed to be myopic. Under this simplifying assumption, \cite{Basar2016-tracking} shows that the principal's optimal information disclosure policy is a set of linear functions of the current state.

To make the problem closer to reality, the authors of \cite{Lingenbrink,Ely-Beep,RENAULT-dynamic,Best1,Best2} consider the myopic assumption only for the information receivers. This set of works studies the interactions between a long-term-optimizing principal and either a myopic receiver or a sequence of short-lived receivers. In the latter case, at each instant of time a new receiver enters the system, forms her belief about the sender's strategy by observing the history of the past messages, takes her action, and then exits the system. In such a case, since the receivers leave the system after taking only one action, they are not concerned about the subsequent effects of their decisions. Therefore, considering a sequence of short-lived receivers is exactly equivalent to assuming that the receiver is myopic. Under this assumption, \cite{Ely-Beep} proposes a generalization of the concavification method for dynamic information design problems. The authors of \cite{RENAULT-dynamic} show that when the receiver is myopic, the greedy disclosure policy where the principal minimizes the amount of information being disclosed in each stage, under the constraint that it maximizes his current payoff, is optimal in many cases, but not always.

There are only a few papers which study the dynamic information design problem with both long-term-optimizing principal and long-term-optimizing receivers \cite{HONRYO-dynamic,Skrzypacz-dynamic,Skrzypacz-selling-info,Ely-goalpost,Ely-sequential-info-design,Basu2017Dynamic,Hung-dynamic,Farhadi2018,Hamid-Allerton,Asu2020}. In \cite{HONRYO-dynamic,Skrzypacz-dynamic,Skrzypacz-selling-info} principal is assumed to have no commitment power. This assumption simplifies the problem as in this case, the principal's policy at each time instant affects the receiver's decisions only at the future and not at the past. Although in some of these works communication is costly, they could be seen, due to lack of commitments, as dynamic versions of cheap-talk problem. The authors in \cite{Ely-goalpost,Ely-sequential-info-design,Basu2017Dynamic,Hung-dynamic,Farhadi2018,Hamid-Allerton,Asu2020} study the dynamic interactions between a long-term-optimizing principal who has full commitment power and a long-term optimizing receiver. In \cite{Ely-goalpost,Ely-sequential-info-design,Basu2017Dynamic,Hung-dynamic} the private information of the principal is considered to be constant and not varying with time. The problem with time-varying private information for the principal is discussed in \cite{Farhadi2018,Hamid-Allerton, Asu2020} for dynamic two-stage settings. The authors of \cite{Asu2020} also tackle the information design problem in an \emph{infinite-horizon} setting. In this setting, they first simplify the problem by restricting attention to a special class of information disclosure mechanisms and then characterize a mechanism that improves the principal's utility, but is not always optimal. 

To the best of our knowledge, this paper is the first work that provides an optimal information disclosure mechanism in a dynamic setting with an arbitrary horizon length and long-term-optimizing principal and receiver, when the principal has full commitment power and his private information evolves over time. 

There are a few works in the literature of information design that combine mechanism design and information design to obtain the benefits of both approaches \cite{Skrzypacz-selling-info,Ely-sequential-info-design}. In these papers, the principal changes both the game-form and the information structure of the play so as to persuade the receiver to take his desirable actions. This approach is not relevant to this paper, as we want to study the effect of the transmitter's information superiority in controlling receivers' behaviors.

\subsection{Contribution} \label{sec:contribution}
We formulate a dynamic information design problem over finite time horizon, where the information transmitter (principal) and the information receiver are long-term optimizing strategic agents. In this problem, the principal wants to use his superior information about system state, which evolves over time, to encourage the receiver to behave such that the interests of principal could be maximized. For this setting, we determine the optimal strategies of both the principal and the receiver. We prove that it is optimal for the principal to give no information to the receiver before a time threshold, run a mixed strategy to confuse the receiver at the threshold time, and reveal the true state afterwards. The principal should announce this information disclosure policy before observing the system state and then stay committed to it forever.

To the best of our knowledge, this is the first instance of a dynamical information design problem with long-term optimizing agents where an explicit solution for the optimal strategies of the agents is determined.

\subsection{Organization of the Paper} \label{sec:outline}
The rest of the paper is organized as follows. We present our
dynamic information design problem with strategic and long-term-optimizing agents in
Section \ref{sec:model}. In Section \ref{sec:problem}, we formulate the principal’s
problem as a dynamic information design problem and discuss its main features. We describe the optimal sequential information disclosure mechanism we propose for the solution to this problem
in Section \ref{sec:solution}. In Section \ref{sec:numerical}, we show the superiority of our proposed mechanism by comparing it to non-strategic mechanisms that are used in real-world applications. We
conclude our paper in Section \ref{sec:conclusion}. The proofs of all the technical results appear in Appendices 1-9.


\section{Model}\label{sec:model}

Consider a Markov chain $\{s_t, t \in \mathcal{T}=\{ 1,2, \ldots, T \}\}$ with state space $\{g (good),\allowbreak b (bad)\}$ and a one-step transition probability matrix 
\begin{equation*}
\mathbb{P} = 
\begin{pmatrix}
1-q & q  \\
0 & 1  
\end{pmatrix}
\end{equation*}
The chain starts in the good state with probability $\mu$, i.e. $P(s_1=g)=\mu$, and then at each instant of time $t = 2, \ldots, T$, may switch from $g$ to $b$ with probability $q$. State $b$ is an absorbing state, meaning that once the Markov chain goes to the bad state, it remains in that state forever. We denote the random time when Markov chain jumps from the good state to the bad state by $\theta$. The situation where the Markov chain starts in the bad state is captured by considering $\theta=1$. If the Markov chain remains in the good state until the end of the time period $T$, we consider $\theta=T+1$. Therefore, the distribution of the random variable $\theta$ is as follows:
\begin{align}\label{eq:theta-dist}
\mathbb{P} (\theta=\theta')=
\begin{cases}
1-\mu,      & \text{if} \hspace{0.15cm} \theta'=1,\\
\mu \hspace{0.05cm} (1-q)^{\theta'-2} \hspace{0.05cm} q,			& \text{if} \hspace{0.15cm} 2 \leq \theta' \leq T,\\
\mu \hspace{0.05cm} (1-q)^{T-1},			& \text{if} \hspace{0.15cm} \theta'=T+1,\\
\end{cases}
\end{align}

There is a strategic detector (she) in the system who wants to detect the jump to the bad state as accurately as possible. Let $\tau$ denote the (random) time the detector declares that the jump has occurred. The detector's cost associated with declaring the jump at time $\tau$ is
\begin{align}\label{eq:cost-detector}
J^D(\tau,\theta)=1_{\{\tau < \theta\}}+1_{\{\tau \geq \theta\}} c(\tau-\theta),
\end{align}
where $1_{\{A\}}$ is the indicator function of an event $A$, which takes value one if $A$ occurs, and zero otherwise. The detector pays one unit of cost if she declares the jump before it actually happens (i.e. false alarm), and pays $c$ units of cost per unit of delayed  detection. The goal of the detector is to choose a detection time $\tau$ so as to minimize the expected value of the cost (\ref{eq:cost-detector}). The detector does not observe the Markov chain's state $s_t$, but she receives some information about it from another agent called the principal.

At each instant of time $t$, the principal (he) observes perfectly the Markov chain's state $s_t$ and sends, according to some information transmission/ disclosure strategy $\boldsymbol{\rho}=(\boldsymbol{\rho}_1,\ldots,\boldsymbol{\rho}_T)$, a message $m_t$ to the detector. When the detector receives the message $m_t$, she updates her belief about the state of the system in a Bayesian way (using $\boldsymbol{\rho}$), and based on her new belief she decides whether or not to declare that the jump has occurred. Let $a_t \in \{k,d\}$ denote the detector's action at time $t \in \mathcal{T}$, where $a_t=k$ indicates that the detector keeps silent at time $t$ and does not declare a jump, and $a_t= d$ indicates that she declares that the jump has occurred. For any fixed choice of the principal's strategy $\boldsymbol{\rho}$, the detector has to solve a quickest detection problem \cite{Shiryaev} to find her best sequence of actions. Therefore, the detector's optimal strategy at each time instant $t$ is described by a threshold; these thresholds are time-varying and depend on the choice of the principal's strategy $\boldsymbol{\rho}$.

The principal's objective is to delay detection of the jump. 
Therefore, utilizing his superior information about the state of the Markov chain, the principal attempts to provide informational incentives to the detector so as to persuade her to keep silent. The principal's utility is
\begin{align}\label{eq:utility-principal}
U^P(\tau)=\tau-1.
\end{align}

The detector's decision depends on her belief about the evolution of the system's unknown state $s_t$; this belief depends on the principal's strategy $\boldsymbol{\rho}$. Therefore, the principle must design a dynamic (over time) information disclosure mechanism in order to influence the evolution of the detector's beliefs and therefore her sequence of actions. 

The above-described model captures fundamental issues arising in strategic information transmission problems, such as deception problems. In deception problems, the information transmitter (he) strategically crafts the information towards his own benefit that is in conflict with the objective of the receiver (she) who uses the information to make decisions which affect the transmitter's utility. In other words, the transmitter, through deceptive signaling, attempts to control the receiver's perception (about the state of nature) so as to lead her to act in line with his own interests. Deception problems arise in Cyber-physical systems, such as power-systems, transportation systems, water-networks, that are vulnerable to adversarial attacks.

In the next section, we formulate this problem as a dynamic information design problem, and design a dynamic information disclosure mechanism that maximizes the principal's utility.

\begin{remark}
	\label{R-model1}
	\normalfont
Our model is similar to Ely's model \cite{Ely-Beep} in that they both consider an uninformed agent (detector) who wants to detect the transition of a two-state Markov chain to its absorbing state. In both models the goal of the information provider is to delay such detection. However, there is a fundamental difference between our model and that of Ely. In Ely's model, the detector uses a time-invariant decision rule that is characterized by a fixed threshold $p^*$. Specifically, the detector declares that the jump from the ``good'' state to the ``bad'' state occurs, at the first time instant $\tau$ at which her (posterior) belief that the Markov chain is in the bad state exceeds $p^*$. Such a decision strategy is not optimal for either a finite-horizon or an infinite-horizon quickest detection problem. In the finite T-horizon quickest detection problem the detector's optimal decision rule is characterized by a sequence of time-varying thresholds that depend on the parameter $c$ (See Eq. (\ref{eq:cost-detector})) and on the functional form of her observations (i.e. the principal's information disclosure strategy). In the infinite-horizon quickest detection problem the detector's optimal decision rule is characterized by a time-invariant threshold as long as the functional form of her observations (i.e. the principal's information disclosure strategy) is time-invariant. It turns out that in our problem as well as in Ely's problem \cite{Ely-Beep} the principal's (optimal) information disclosure strategy is not time-invariant, therefore, the detector's optimal decision strategy is not characterized by a time-invariant threshold.

In our problem the detector's decision rule is the optimal decision rule for the T-horizon quickest detection problem where the functional form of her observations is the principal's optimal information disclosure strategy.
\end{remark}

\section{The Dynamic Information Design Problem} \label{sec:problem}

\subsection{Problem Formulation} \label{sec:problem-formula}

A dynamic information disclosure mechanism specifies the set of messages $\mathcal{M}_t$ that the principal sends to the detector at each instant of time $t\in \mathcal{T}$, along with a distribution over $\mathcal{M}_t$ given all the information available to the principal at time $t$. The principal's information at time $t$ consists of 1) the history of evolution of the state (which the principal observes perfectly) up to time $t$, i.e. $s_{1:t}$, 2) the history of his past messages to the detector, i.e. $m_{1:t-1}$, and 3) the history of the detector's past actions, i.e. $a_{1:t-1}$. The set of dynamic information disclosure mechanisms is completely general, it includes the extremes of full information, no information, and all conceivable intermediate mechanisms. The full information mechanism is the rule that reveals perfectly the current state $s_t$ : i.e., $\mathcal{M}_t=\{g,b\}$ and $\mathbb{P} (m_t=s_t)=1$. A no information mechanism is obtained when the set of messages $\mathcal{M}_t$ has only one element and the principle sends that single message irrespective of the system state.

As it is clear from the above examples, a dynamic information disclosure mechanism could be very complicated since there is no restriction on the set of messages $\mathcal{M}_t$ or the probability distribution on $\mathcal{M}_t$, $t \in \mathcal{T}$. However, it is shown in \cite{Bayesian-persuasion} that there is no loss of generality in restricting attention to \emph{direct} dynamic information disclosure mechanisms that are \emph{obedient}. Thus, we concentrate on direct dynamic information disclosure mechanisms that are obedient.

In a direct information disclosure mechanism, at each instant of time, the principal directly recommends to the detector the action she should take. In our problem, the detector's possible actions at each time $t$ are to either keep silent ($a_t=k$) or declare the jump ($a_t=d$). Therefore, the set of messages used by the principal at each time $t$ is $\mathcal{M}_t=\mathcal{M}=\{k,d\}$, where $k$ is a recommendation to keep silent, and $d$ is a recommendation to declare a jump. As a result, the principal's behavior in a direct information disclosure mechanism can be described by a recommendation policy $\boldsymbol{\rho}=(\rho_{t}^{s_{1:t},m_{1:t-1},a_{1:t-1}}, t \in \mathcal{T})$, where $\rho_{t}^{s_{1:t},m_{1:t-1},a_{1:t-1}}$ is the probability according to which the principal sends message $k$ to the detector (i.e. recommends her to keep silent), when the sequence of the states he has observed up to time $t$ is $s_{1:t}=(s_1,\ldots,s_t)$, the history of the past messages he has sent is $m_{1:t-1}=(m_1,\ldots,m_{t-1})$, and the history of the detector's past actions is $a_{1:t-1}=(a_1,\ldots,a_{t-1})$. For each $t \in \mathcal{T}$, $s_{1:t} \in \{g,b\}^{t}$, and $m_{1:t-1}, a_{1:t-1} \in \{k,d\}^{t-1}$, the principal sends message $d$ to the detector with probability $1-\rho_{t}^{s_{1:t},m_{1:t-1},a_{1:t-1}}$. There are two features in our problem that help us to simplify the principal's information.

1) At each instant of time, the detector has two actions one of which (i.e. declaring the jump) terminates the whole process. Therefore, making a recommendation at time $t$ is meaningful for the principal only if the detector kept silent at all the previous times, i.e. $a_{1:t-1}=(k)^{t-1}$. Thus, $a_{1:t-1}$ can be omitted from the principal's information.

2) The bad state of our Markov chain is an absorbing state. Therefore, the state evolution of the Markov chain until time $t$ is of the form $s_{1:t}=((g)^{\theta_t-1},(b)^{t-\theta_t+1})$, where $\theta_t$ could take any integer value between $1$ and $t+1$. We define $\theta_t$ as the earliest possible time for the jump based on the principal's information at time $t$; i.e. $\theta_t=\min{(\theta,t+1)}$.
The parameter $\theta_t$ is equal to $\theta$ if the jump has occurred up to time $t$, however it takes value $t+1$ when the principal finds the Markov chain in the good state at time $t$. Because of this feature, we can represent the recommendation policy $\rho_{t}^{s_{1:t},m_{1:t-1}}$ of each time $t$ by $\rho_{t}^{\theta_t,m_{1:t-1}}$, where $\theta_t \in  \{1, \ldots, t+1 \}$.

When the principal designs an information disclosure mechanism, he announces the corresponding recommendation policy $\boldsymbol{\rho}$ to the detector and commits to it.

The detector is strategic; she utilizes the information she receives from the principal to her own advantage and does not necessarily follow
the actions recommended by the principal. Therefore, to achieve his goal, the principal must design an information disclosure mechanism that possesses the obedience property, that is, it provides the detector with strong enough incentives to follow his recommendations.

At each time $t \in \mathcal{T}$, the long-term-optimizing detector obeys the principal if the recommended action $m_t$ minimizes her conditional expected continuation cost, i.e.,
\begin{align}\label{eq:obedience-constraint}
m_t=\argmin_{a_t}{\left[\min_{\gamma_{t+1:T}}{\mathbb{E}^{\boldsymbol{\rho}}_{t:T} \{J^D(\tau,\theta) | m_{1:t}\}}\right]}, \hspace{0.1cm} \forall t, m_{1:t},
\end{align}
where $\gamma_{t+1:T}$ denotes her decision strategy profile from time $t+1$ up to $T$, and $\min_{\gamma_{t+1:T}}{\mathbb{E}^{\boldsymbol{\rho}}_{t:T} \{J^D(\tau,\theta) | m_{1:t}\}}$ is her minimum expected continuation cost
conditional on her information $m_{1:t}$ when the principal commits to the information disclosure strategy/mechanism $\boldsymbol{\rho}$. Therefore, to check the obedience property of a direct information disclosure mechanism at any time $t$, $t=1,2,\ldots,T$, we need to solve the series/sequence of optimization problems 
\begin{align}\label{eq:obedience-constraintt}
\min_{a_t}{\left[\min_{\gamma_{t+1:T}}{\mathbb{E}^{\boldsymbol{\rho}}_{t:T} \{J^D(\tau,\theta) | m_{1:t}\}}\right]},
\end{align}
for all $m_{1:t}$. Solving these optimization problems is a challenging and time-consuming task. However, the one-shot deviation principle allows us to derive the obedience constraints by assuming that the detector sticks to the obedient strategy in the future. Specifically, considering a fixed direct information disclosure mechanism, the whole process can be seen as a finite extensive-form game between the detector and nature (principal), where at each stage $t$ nature sends a signal $m_t$ to the detector according to $\rho_t$ and then the detector takes an action $a_t$. With this interpretation, the set of obedience constraints (\ref{eq:obedience-constraint}) is a necessary and sufficient set of conditions for the strategy profile $a_t=m_t$, for $t \in \mathcal{T}$, to be a subgame-perfect Nash equilibrium (SPNE) \cite{Myerson1991}. According to the one-shot deviation principle a strategy profile $\gamma$ is a SPNE if and only if no player can increase their payoffs by deviating from $\gamma$ for one period and then reverting to the strategy \cite{fudenberg1991}. Therefore, a simpler version of the obedience constraints is as follows:
\begin{align}\label{eq:obedience-constraint2}
m_t\hspace{-0.06cm}=\hspace{-0.06cm}\argmin_{a_t}{\hspace{-0.06cm}\left[\mathbb{E}^{\boldsymbol{\rho}}_{t:T} \{J^D(\tau,\theta) | m_{1:t}, a_{t+1:T}\hspace{-0.06cm}=\hspace{-0.06cm}m^{\boldsymbol{\rho}}_{t+1:T}\}\right]},
\end{align}
for all $t$ and $m_{1:t}$, where the condition $a_{t+1:T}=m^{\boldsymbol{\rho}}_{t+1:T}$ of the expectation reflects the fact that the detector obeys the recommendations from time $t+1$ onward, and the notation $m^{\boldsymbol{\rho}}_{t+1:T}$ indicates that the messages $m_{t+1:T}$ are generated according to the policy $\boldsymbol{\rho}$.

Equation (\ref{eq:obedience-constraint2}) includes $\sum_{t=1}^T{2^t}=2(2^T-1)$ constraints. These constraints force the detector to obey the recommendations after each message history $m_{1:t}$. However, due to the nature of our problem, the first time the detector is recommended to declare a jump in an obedient mechanism, she will do so and the process will terminate. Therefore, the message histories with at least one $d$ in the past are not going to occur and need not be checked. This feature reduces the obedience constraints need to be considered to the following:
\begin{align}\label{eq:obedience-constraint-k}
\begin{split}
&\mathbb{E}^{\boldsymbol{\rho}}_{t:T} \{J^D(\tau,\theta) | m_{1:t}=(k)^t, a_t=k, a_{t+1:T}=m^{\boldsymbol{\rho}}_{t+1:T}\} \\ 
&\leq \mathbb{E}^{\boldsymbol{\rho}}_{t:T} \{J^D(\tau,\theta) | m_{1:t}=(k)^t, a_t=d\}, \forall t \in \mathcal{T},
\end{split}
\end{align}
\begin{align}\label{eq:obedience-constraint-d}
\begin{split}
&\mathbb{E}^{\boldsymbol{\rho}}_{t:T} \{J^D(\tau,\theta) | m_{1:t}\hspace{-0.06cm}=\hspace{-0.06cm}((k)^{t-1},d), a_t\hspace{-0.06cm}=\hspace{-0.06cm}k, a_{t+1:T}\hspace{-0.06cm}=\hspace{-0.07cm}m^{\boldsymbol{\rho}}_{t+1:T}\} \\
&\geq \mathbb{E}^{\boldsymbol{\rho}}_{t:T} \{J^D(\tau,\theta) | m_{1:t}\hspace{-0.07cm}=\hspace{-0.07cm}((k)^{t-1},d), a_t\hspace{-0.07cm}=\hspace{-0.07cm}d\}, \forall t \hspace{-0.07cm} \in \hspace{-0.07cm} \mathcal{T},
\end{split}
\end{align}
($2T$ constraints).

Now that we derived the obedience constraints, we go through calculating the principal's utility. When the detector follows the recommendation policy $\boldsymbol{\rho}$, the expected utility the principal gets is
\begin{align}\label{eq:principal-utility}
\begin{split}
&\mathbb{E}^{\boldsymbol{\rho}} \{U^P(\tau)\}=\mathbb{E}^{\boldsymbol{\rho}} \{\tau-1\}=\mathbb{E}^{\boldsymbol{\rho}}\{\sum_{t=1}^T{1_{\{a_{1:t} = (k)^t\}}}\} \\
&=\sum_{t=1}^T{\mathbb{P}(a_{1:t} = (k)^t| \boldsymbol{\rho})}=\sum_{t=1}^T{\mathbb{P}(m^{\boldsymbol{\rho}}_{1:t} = (k)^t)}\\
&=\sum_{t=1}^T{\sum_{\theta'=1}^{T+1}{\mathbb{P}(\theta=\theta') \prod_{t'=1}^t{\rho_{t'}^{\min{(\theta',t'+1)},(k)^{t'-1}}}}}.
\end{split}
\end{align}
Therefore, we can formulate the information design problem (Problem \ref{eq:info-design-problem}) for the principal as follows:
\begin{align} \tag{\textbf{P}} \label{eq:info-design-problem}
&\max_{\boldsymbol{\rho}}{\mathbb{E}^{\boldsymbol{\rho}} \{U^P(\tau)\}},\nonumber\\
\textit{subject to}  & \textit{ the obedience constraints (\ref{eq:obedience-constraint-k})-(\ref{eq:obedience-constraint-d}),} \nonumber\\
& 0 \leq \rho_{t}^{\theta_t,m_{1:t-1}} \leq 1, \hspace{0.1cm} \forall t, \theta_t, m_{1:t-1}. \nonumber
\end{align}
That is, the principal wants to choose a feasible dynamic information disclosure mechanism $\boldsymbol{\rho}$ that satisfies the obedience constraints (\ref{eq:obedience-constraint-k})-(\ref{eq:obedience-constraint-d}) and maximizes his expected utility given by (\ref{eq:principal-utility}).

\subsection{Features of The Problem} \label{sec:problem-features}

Solving the optimization problem (\ref{eq:info-design-problem}) is a formidable task as the optimization variables are strongly coupled with many non-convex constraints. This strong coupling can be seen by taking a closer look at the expectations appearing in the obedience constraints (\ref{eq:obedience-constraint-k})-(\ref{eq:obedience-constraint-d}). According to (\ref{eq:cost-detector}), the detector's expected continuation cost from time $t$ onward when she has received messages $m_{1:t}$ and decides to declare the jump at time $t$ is
\begin{align}\label{eq:cost-declare}
\mathbb{E}^{\boldsymbol{\rho}}_{t:T} \{J^D(\tau,\theta) | m_{1:t}, a_t=d\}= \mathbb{P} (s_t=g | m_{1:t}),
\end{align}
i.e., the probability of false alarm at $t$. If the detector decides to keep silent at time $t$ and sticks to the obedient strategy at the future, her expected continuation cost is
\begin{align}\label{eq:cost-silence}
&\mathbb{E}^{\boldsymbol{\rho}}_{t:T} \{J^D(\tau,\theta) | m_{1:t}, a_t=k, a_{t+1:T}=m^{\boldsymbol{\rho}}_{t+1:T}\}= \nonumber\\
& c (1\hspace{-0.07cm}-\hspace{-0.07cm}\mathbb{P} (s_t\hspace{-0.07cm}=\hspace{-0.07cm}g | m_{1:t})) + \nonumber\\
&\mathbb{E}^{\boldsymbol{\rho}}_{t+1:T} \{J^D(\tau,\theta) | m_{1:t}, a_t\hspace{-0.07cm}=\hspace{-0.07cm}k, a_{t+1:T}\hspace{-0.07cm}=\hspace{-0.07cm}m^{\boldsymbol{\rho}}_{t+1:T}\},
\end{align}
where the first term on the right hand side is the expected delay cost at time $t$, and the second term denotes the expected value of all the future costs from time $t+1$ onward. Substituting (\ref{eq:cost-declare})-(\ref{eq:cost-silence}) in (\ref{eq:obedience-constraint-k})-(\ref{eq:obedience-constraint-d}) shows that at each time $t$, the detector's decision to obey or disobey the principal's recommendation depends on two factors: 

(i) the belief the detector has about the good state of the Markov chain, i.e. $\mathbb{P} (s_t=g| m_{1:t})$; this belief is constructed according to Bayes' rule from the past messages $m_{1:t}$ she received, hence the past recommendation rules $\rho_{t'}^{\theta_{t'},m_{1:t'-1}}$ ($t' \leq t$) that generate these messages, as follows:
\small
\begin{align}\label{eq:current-belief}
&\mathbb{P} (s_t=g | m_{1:t})=\frac{\mathbb{P}(s_t=g,m_{1:t})}{\mathbb{P} (m_{1:t})}=\frac{\mathbb{P}(s_t=g) \mathbb{P} (m_{1:t} | s_t=g)}{\sum_{\theta'=1}^{T+1}{\mathbb{P} (\theta=\theta')\mathbb{P} (m_{1:t} | \theta=\theta')}}=\nonumber\\ 
&\frac{\mathbb{P}(\theta \hspace{-0.07cm}>\hspace{-0.07cm} t) \prod_{t'\hspace{-0.03cm}=\hspace{-0.03cm}1}^t{[\rho_{t'}^{t'\hspace{-0.04cm}+\hspace{-0.04cm}1,m_{1:t'\hspace{-0.04cm}-\hspace{-0.04cm}1}}1_{\{m_{t'} = k\}}+(1-\rho_{t'}^{t'+1,m_{1:t'-1}})1_{\{m_{t'} = d\}}]}}{\sum_{\theta'\hspace{-0.03cm}=\hspace{-0.03cm}1}^{T\hspace{-0.03cm}+\hspace{-0.03cm}1}{\mathbb{P} (\theta\hspace{-0.05cm}=\hspace{-0.05cm}\theta')\prod_{t'\hspace{-0.05cm}=\hspace{-0.05cm}1}^{t}{[\rho_{t'}^{\min{(\theta',t'+1)},m_{1:t'\hspace{-0.05cm}-\hspace{-0.05cm}1}}1_{\{m_{t'} = k\}}\hspace{-0.07cm}+\hspace{-0.07cm}(1\hspace{-0.07cm}-\hspace{-0.07cm}\rho_{t'}^{\min{(\theta',t'+1)},m_{1:t'\hspace{-0.05cm}-\hspace{-0.05cm}1}})1_{\{m_{t'} = d\}}]}}}.
\end{align}
 \normalsize
(ii) the cost the detector expects to incur in the future if she remains silent at time $t$ and follows the recommendations afterwards, i.e. $\mathbb{E}^{\boldsymbol{\rho}}_{t+1:T} \{J^D(\tau,\theta) | m_{1:t},\allowbreak a_t=k, a_{t+1:T}=m^{\boldsymbol{\rho}}_{t+1:T}\}$. This expected cost depends on the messages the detector expects to receive in the future. These messages depend on the future recommendation rules to which the principal commits, i.e. $\rho_{t'}^{\theta_{t'},m_{1:t'-1}}$, $t' > t$, $\forall \theta$.

Therefore, the detector's decision at each time $t$ depends not only on the recommendation policy for time $t$, but also on the recommendation policies for all times before and after $t$. The dependence of the detector's decision at each time $t$ on the recommendation policy for the entire horizon makes the discovery of an optimal information disclosure mechanism that satisfies the obedience property very challenging. This is mainly because any change in the principal's recommendation policy at any time $t$ affects the detector's decisions at all times. Therefore, the principal cannot optimize the recommendation policies of different time slots separately just by considering the obedience constraints at that time, instead he needs to optimize the recommendation policies of the whole horizon simultaneously.

\section{An Optimal Sequential Information Disclosure Mechanism}\label{sec:solution}

In this section we present the main result of the paper, namely, a dynamic information disclosure mechanism that solves the principal's problem, expressed by Problem (\ref{eq:info-design-problem}). The mechanism is described in Theorem \ref{T-optimal} below. To state Theorem \ref{T-optimal} we need the following definitions.

\begin{definition}
	\label{class-state-only}
	\normalfont
Define by $\Gamma^M$ the class of information disclosure mechanisms $\boldsymbol{\rho}=(\rho_{t}^{s_{t},m_{1:t-1}}, t \in \mathcal{T})$ where $\rho_{t}^{s_{t},m_{1:t-1}}$ depends on the message profile $m_{1:t-1}$ received by the detector up to time $t-1$ and only the current state $s_{t}$ of the Markov chain at time $t$ (not on the state evolution $s_{1:t}$).
\end{definition}

\begin{definition}
	\label{time-based-prioritized}
	\normalfont
A mechanism $\boldsymbol{\rho}=(\rho_{t}^{s_{t},m_{1:t-1}}, t \in \mathcal{T}) \in \Gamma^M$ is called \emph{time-based prioritized} if:
\begin{enumerate}
\item $\rho_{t}^{g,m_{1:t-1}}=1$ for all $t$ and all $m_{1:t-1}$;
\item  there is no time $t$, $t \in \mathcal{T}$, such that $\rho_{t+1}^{b,(k)^{t}} >0$ while $\rho_t^{b,(k)^{t-1}} <1$, where $(k)^{t}=(k,\ldots,k)$ is a vector of length $t$ with all components equal to $k$; and
\item for all $t \in \mathcal{T}$, $\rho_t^{b,m_{1:t-1}}$ is arbitrary, $0 \leq \rho_t^{b,m_{1:t-1}} \leq 1$, when $m_{1:t-1} \neq (k)^{t}$.
\end{enumerate}
\end{definition}
We denote by $\Gamma^{MP}$ the class of time-based prioritized mechanisms. In information disclosure mechanism $\boldsymbol{\rho} \in \Gamma^{MP}$, the priority of keeping the detector silent at each time $t > \theta$ is higher than keeping her silent at $t+1$. Therefore, if $s_t=b$ the principal does not put any effort in manipulating the detector's information at $t+1$ unless there is no room for improving his own performance at time $t$. As a consequence of Definition \ref{time-based-prioritized}, for each mechanism $\boldsymbol{\rho} \in \Gamma^{MP}$, there is a threshold $n_p$ such that
\begin{align}\label{eq:time-pr-mech}
\rho_t^{b,(k)^{t-1}}=
\begin{cases}
1,     & t < n_p,\\
0,     & t > n_p,\\
q_{n_p},  & t=n_p, q_{n_p} \in \left[0,1\right].
\end{cases}
\end{align}
Therefore, any $\boldsymbol{\rho} \in \Gamma^{MP}$ can be uniquely described by $\boldsymbol{\rho}=(n_p,q_{n_p})$.
\begin{theorem}
	\label{T-optimal}
	\normalfont
Without loss of optimality, in Problem (\ref{eq:info-design-problem}) the principal can restrict attention to time-based prioritized mechanisms. Determining an optimal time-based prioritized mechanism $\boldsymbol{\rho}^*$ for Problem (\ref{eq:info-design-problem}) is equivalent to finding $n_p^*+q_{n_p}^*=max{\left\{n_p+q_{n_p}\right\}}$ such that the mechanism $(n_p,q_{n_p})$ satisfies all the obedience constraints. The optimal time-based prioritized information disclosure mechanism can be obtained by Algorithm 1 above.

\begin{algorithm}[t]\label{alg1}
\For{$n_p=1,\ldots,T$}{
 \small $q_{n_p} \leftarrow \min_{t \leq n_p}{\hspace{-0.07cm}\left\{\frac{1}{\mathbb{P}(\theta \leq n_p)}(\frac{\mathbb{P}(\theta > t)}{c}\hspace{-0.07cm}-\hspace{-0.1cm}\sum_{l=t}^{n_p\hspace{-0.07cm}-\hspace{-0.03cm}1}{\mathbb{P}(\theta \hspace{-0.04cm} \leq \hspace{-0.04cm} l)})\right\}}$ \;
\normalsize
\If{$q_{n_p} < 1$}{
break\;
}
}
$n_p^* \longleftarrow n_p$ \;
$q_{n_p}^* \longleftarrow q_{n_p}$.
 \caption{An optimal time-based prioritized information disclosure mechanism}
\end{algorithm}

\noindent The principal's expected utility at $\boldsymbol{\rho}^*=(n_p^*,q_{n_p}^*)$ is 
\begin{align}\label{eq:utility-time-pr-opt}
\mathbb{E}^{\boldsymbol{\rho}^*} \{U^P\}= n_p^*-1+\mathbb{P}(\theta \leq n_p^*) q_{n_p}^*+\sum_{t=n_p^*}^T{\mathbb{P}(\theta > t)}.
\end{align}
\end{theorem}

\noindent \textbf{Outline of the proof of Theorem \ref{T-optimal}}

Theorem \ref{T-optimal} provides us with an optimal sequential information disclosure mechanism. We prove this theorem in three steps. 

In the first step, we reduce the complexity of the optimization problem (\ref{eq:info-design-problem}) by reducing/simplifying the domain of $\boldsymbol{\rho}$. To this end, we show that:

1) the recommendation policy the principal uses at any time $t$ when he has advised the detector to declare the jump at least once before $t$, plays no role in problem (\ref{eq:info-design-problem});

2) without loss of optimality, the principal can restrict attention to mechanisms $\boldsymbol{\rho}=(\rho_{t}^{s_{t},m_{1:t-1}}, t \in \mathcal{T})\in \Gamma^M$. In this class of mechanisms, the recommendation policy at any time $t$ depends only on the state $s_{t}$ of the Markov chain and the principal's previous messages $m_{1:t-1}$ (not on the state evolution $s_{1:t-1}$).

In the second step, we prove that $\rho_{t}^{g,m_{1:t-1}}=1$, for $t \in \mathcal{T}$ and all $m_{1:t-1}$. Thus, when the Markov chain is in the good state, the principal always recommends the detector to wait.

In the third step, we use the results of the first two steps to prove that restricting attention to the class of  time-based prioritized mechanisms is without loss of optimality. Then, we determine an optimal solution for the dynamic information disclosure problem (\ref{eq:info-design-problem}) in this class.

\textbf{Proof of Theorem \ref{T-optimal}:} We prove in the appendix the lemmas appearing in each step.

\textbf{Step 1.} As we discussed in Section \ref{sec:problem}, the principal's behavior in a direct dynamic information disclosure mechanism is described by a recommendation policy $\boldsymbol{\rho}=(\rho_{t}^{\theta_t,m_{1:t-1}}, t \in \mathcal{T})$, where $\rho_{t}^{\theta_t,m_{1:t-1}}$ is the probability with which the principal recommends the detector to keep silent at time $t$, when $\theta_t=\min{(\theta,t+1)}$ and the message profile that has been sent so far is $m_{1:t-1}$. For each time $t$, $\theta_t$ could take any integer value between $1$ and $t+1$. Moreover, the principal's message may take two values at each time slot, so we can have $2^{t-1}$ different message profiles at each time $t$. Therefore, to design an optimal direct dynamic information disclosure mechanism we need to determine the optimal values of $\sum_{t=1}^T{2^{t-1}(t+1)}=2^T T$ different variables. This number grows exponentially with the horizon length $T$. Furthermore, these variables are coupled with one another through the obedience constraints (\ref{eq:obedience-constraint-k})-(\ref{eq:obedience-constraint-d}), hence their optimal values must be determined simultaneously. The large number of design variables and their dependence on one another make the determination of an optimal information disclosure mechanism a formidable task. Therefore, our goal in this step is to reduce the number of design variables, without any loss of optimality. We achieve our goal via the results of Lemmas \ref{L-message-history} and \ref{L-state-history}. Before stating Lemmas \ref{L-message-history} and \ref{L-state-history}, we define the following function that counts the number of time epochs the detector was recommended to declare the jump in the past.
\begin{definition}
	\label{No-d-message}
	\normalfont
For each time $t$ and each message history $m_{1:t-1}$, we define $\mathcal{N}_{d}(m_{1:t-1})$ as the number of time slots where message $d$ has been sent to the detector.
\end{definition}

\begin{lemma}
	\label{L-message-history}
	\normalfont
For each time $t$, the recommendation policy $\rho_{t}^{\theta_t,m_{1:t-1}}$ needs to be designed only for message profiles $m_{1:t-1}$ with $\mathcal{N}_{d}(m_{1:t-1}) \leq 1$. The part of the recommendation policy related to cases where the message $d$ has been sent more than once has no effect on either the obedience property of the mechanism or the utility it gets to the principal.
\end{lemma}

As a result of Lemma \ref{L-message-history}, an optimal recommendation policy must determine the optimal values of $\sum_{t=1}^T{t(t+1)}=\frac{1}{3} T (T+1)(T+2)$ variables. This number grows polynomially rather than exponentially with the horizon length $T$, so the complexity of the information disclosure problem is significantly reduced.

\begin{lemma}
	\label{L-state-history}
	\normalfont
Without loss of optimality, for each message profile $m_{1:t-1}$, the principal can restrict attention to recommendation policies that depend, at each time $t$, on the current state $s_t$ of the Markov chain and not on the exact time the jump has occurred.
\end{lemma}

As a result of Lemma \ref{L-state-history}, at each time $t$, and for each message profile $m_{1:t-1}$, the principal needs to consider only two recommendation strategies, one when $s_t=g$ the other when $s_t=b$. Therefore, the total number of design variables is $\sum_{t=1}^T{2t}=T(T+1)$, which grows as $T^2$ rather than $T^3$.

\textbf{Step 2.} We derive an optimal recommendation strategy for the principal when the Markov chain is in the good state. 

\begin{lemma}
	\label{L-g-state}
	\normalfont
If at any time $t$ the Markov chain is in the good state, irrespective of the message profile $m_{1:t-1}$, it is always optimal for the principal to recommend the detector to keep silent. That is 
\begin{align}\label{eq:rho-good}
\rho_{t}^{* \hspace{0.05cm} g,m_{1:t-1}}=1, \forall m_{1:t-1}, \forall t \in \mathcal{T},
\end{align}
\end{lemma}
The result of Lemma \ref{L-g-state} is intuitive, because when the state of the Markov chain is good there is no conflict of interest between the principal and the detector. In this state, the principal wants to prevent the detector from declaring a jump, and the detector herself has no incentive to create a false alarm. Therefore, there is no incentive for the principal to mislead the detector.

As a result Lemma \ref{L-g-state}, when the detector is recommended to declare a jump, she is absolutely sure that the Markov chain is in the bad state, thus, she declares a jump. Therefore, the obedience constraints (\ref{eq:obedience-constraint-d}) corresponding to situations where the detector receives recommendation $m_t=d$ are automatically satisfied and can be neglected in the rest of the design process. Moreover, if any message $d$ has been sent by the principal in the past, the detector declares a jump right after receiving $d$ and the whole process terminates. Therefore, we do not need to design a recommendation policy for message profiles that contain at least one $d$. Consequently, the only variable we should design at any time $t$, is the probability of recommending the detector to declare a jump when $s_t=b$ and $m_{1:t-1}=(k)^{t-1}$. Finding the optimal values of these variables is the subject of the next step.

\textbf{Step 3.} First, we show that, without loss of optimality, the principal can restrict attention to time-based prioritized mechanisms. Then, we determine such optimal mechanism.

\begin{lemma}
	\label{L-priority}
	\normalfont
Without loss of optimality, the principal can restrict attention to the class of time-based prioritized mechanisms.
\end{lemma}

We proceed now to complete the proof of our main result (Theorem \ref{T-optimal}). The expected utility of a principal who uses time-based prioritized mechanism $\boldsymbol{\rho}=(n_p,q_{n_p})$ is
\begin{align}\label{eq:utility-time-pr}
\mathbb{E}^{(n_p,q_{n_p})} \{U^P\}\hspace{-0.07cm}= \hspace{-0.07cm} n_p\hspace{-0.07cm} -\hspace{-0.07cm}1\hspace{-0.07cm}+\hspace{-0.07cm}\mathbb{P}(\theta \leq n_p)q_{n_p}\hspace{-0.1cm}+\hspace{-0.1cm}\sum_{t=n_p}^T{\mathbb{P}(\theta \hspace{-0.04cm} > \hspace{-0.04cm} t)}.
\end{align}
This can be seen as follows. Before the threshold time is reached, the detector is always recommended to keep silent. Therefore, at the first $n_p-1$ time slots silence is guaranteed. At the threshold period $n_p$, if the Markov chain is in the bad state (prob. $\mathbb{P}(\theta \leq n_p)$), the detector remains silent with probability $q_{n_p}$. However, if the Markov chain is in the good state (prob. $\mathbb{P}(\theta > n_p)$), the detector remains silent for sure. At each time $t$ after the threshold period, the detector keeps quiet only if the jump has not occurred. The probability of this event is $\mathbb{P}(\theta > t)$.

\begin{figure}[t]
\centering
\includegraphics[width=0.7 \textwidth, height=5cm]{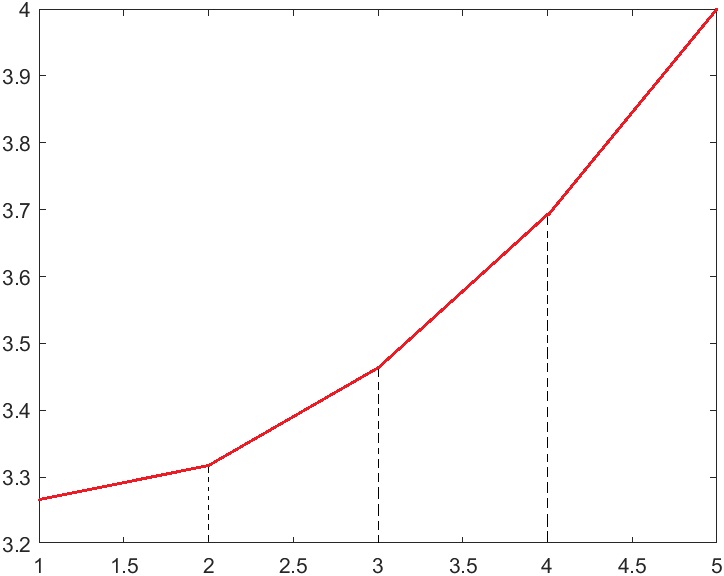}
\caption{An example of the Principal's utility function in a time-based prioritized mechanism}
\label{fig:utility-time-based}
\end{figure}

Equation (\ref{eq:utility-time-pr}) shows that for each constant threshold $n_p$, the principal's expected utility is an increasing linear function of $q_{n_p}$. Moreover, we have $\mathbb{E}^{(n_p,1)} \{U^P\}=\mathbb{E}^{(n_p+1,0)} \{U^P\}$. This is true, simply because $\boldsymbol{\rho}=(n_p,1)$ and $\boldsymbol{\rho'}=(n_p+1,0)$ are actually two representations of the same mechanism. We can easily conclude from these two facts that the principal's expected utility is a piecewise linear function of $n_p+q_{n_p}$ as depicted in Fig. \ref{fig:utility-time-based}. The slope of the segments increases whenever the variable $n_p+q_{n_p}$ takes an integer value.

The arguments above show that finding the optimal time-based prioritized mechanism is equivalent to finding the maximum value of $n_p\hspace{-0.04cm}+\hspace{-0.04cm}q_{n_p}$ such that the mechanism $\boldsymbol{\rho}\hspace{-0.07cm}=\hspace{-0.07cm}(n_p,q_{n_p})$ satisfies the obedience constraints. With some algebra, it can be shown that for a time-based prioritized mechanism $\boldsymbol{\rho}$ the obedience constraints (\ref{eq:obedience-constraint-k}) can be simplified to
\begin{align}\label{eq:obed-time-pr}
c(\sum_{l=t}^{n_p\hspace{-0.03cm}-\hspace{-0.03cm}1}{\mathbb{P}(\theta \hspace{-0.07cm} \leq \hspace{-0.07cm} l)} \hspace{-0.07cm} +\hspace{-0.07cm}  \mathbb{P}(\theta \leq n_p) q_{n_p}) \leq \mathbb{P}(\theta > t), \forall t \leq n_p.
\end{align}
These constraints are very intuitive as for each time $t$, 1) the right-hand side of (\ref{eq:obed-time-pr}) is the expected cost of declaring the jump at time $t$; and 2) the left-hand side of (\ref{eq:obed-time-pr}) is the expected continuation cost that the detector incurs from time $t$ onward, when she follows the recommendations made by the mechanism $\boldsymbol{\rho}=(n_p,q_{n_p})$. Therefore, constraints (\ref{eq:obed-time-pr}) simply say that a time-based prioritized mechanism is obedient if and only if the detector finds declaring the jump more costly than keeping silent at each time $t \leq n_p$ when she is recommended to stay quiet.

The left-hand side of each obedience constraint in (\ref{eq:obed-time-pr}) is increasing in terms of $q_{n_p}$. Therefore, if the obedience conditions are satisfied for a mechanism $\boldsymbol{\rho}=(n_p,q_{n_p})$, they are also satisfied for mechanisms with smaller values for $q_{n_p}$. Given that the mechanisms $(n_p,0)$ and $(n_p-1,1)$ are the same, we can conclude that obedience of a time-based-prioritized mechanism $\boldsymbol{\rho}=(n_p,q_{n_p})$ implies the obedience of all time-based-prioritized mechanisms with smaller values of $n_p+q_{n_p}$. Therefore, there is a threshold $k^*$ such that the set of all obedient time-based-prioritized mechanisms consists of the mechanisms for which $n_p+q_{n_p}$ takes a value smaller than the threshold $k^*$. The fact that the principal's expected utility is increasing in terms of $n_p+q_{n_p}$ implies that the mechanism with $n_p+q_{n_p}=k^*$ is optimal. This optimal mechanism is uniquely determined by Algorithm 1.



\begin{figure}[t]
\centering
\includegraphics[width= \textwidth, height=5cm]{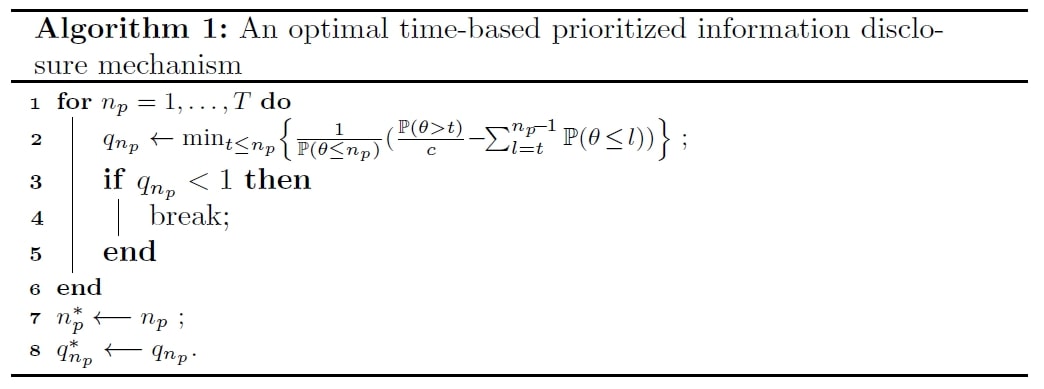}
\end{figure}

Algorithm 1 works as follows: It iterates over $n_p=1,\ldots,T$ and at each iteration it computes the maximum value of $q_{n_p}$ such that $\boldsymbol{\rho}=(n_p,q_{n_p})$ satisfies the obedience constraints (\ref{eq:obed-time-pr}) for all $t \leq n_p$. Achieving a maximum greater than $1$ means that the mechanism $(n_p,1)=(n_p+1,0)$ not only satisfies all the obedience constraints, but also has no binding constraints. This means that there is still more room for improvement. Therefore, the algorithm goes to the next iteration to find a mechanism with the greater utility for the principal. If at some iteration $n_p^*$ we obtain a maximum of less than one for $q_{n_p}$ we stop. The mechanism $(n_p^*,q_{n_p}^*)$ satisfies all obedience constraints and there are binding obedience constraints. Therefore, $k^*=n_p^*+q_{n_p}^*$ is the optimal threshold which cannot be enhanced anymore anymore, hence $\boldsymbol{\rho}^*=(n_p^*,q_{n_p}^*)$ is an optimal information disclosure mechanism.

The proof of Theorem \ref{T-optimal} is now complete.

\section{Results and Discussion} \label{sec:numerical}

In this section, we discuss and highlight some interesting features of our designed optimal mechanism obtained with Algorithm 1. We also run some numerical experiments to observe its performance. 

\textbf{Feature 1}. The mechanism is almost independent of the horizon length $T$. The only effect of the horizon length on the optimal mechanism is to limit the threshold $n_p$ from above. Therefore, as long as the optimal threshold $n_p^{*}$ is an interior point, changing the time horizon $T$ does not change the optimal mechanism.

\textbf{Feature 2}. The optimal mechanism $\boldsymbol{\rho}^*=(n_p^*,q_{n_p}^*)$ we propose is a time-based prioritized mechanism. The principal in a time-based prioritized mechanism employs three different strategies in different time regions (See Eq. (\ref{eq:time-pr-mech})). 

\textbf{Region 1:} In the first region which consists of times before the threshold $n_p^{*}$, irrespective of the Markov chain's state $s_t$, the principal recommends the detector to keep silent. During this time interval, the principal's messages are independent of the Markov chain's state. These messages give no information to the detector and hence, without loss of optimality, can be removed from the mechanism. Therefore, this region can be referred to as the \emph{no-information region}.

\textbf{Region 2:} In the second region which takes only one time slot (i.e. $t=n_p^{*}$), the principal runs a mixed/randomized strategy to hide his information. In this time slot, the principal always recommends the detector to keep silent if the Markov chain is in the good state. However, when the Markov chain is in the bad state, he reveals this undesirable news only with a probability of $q_{n_p}^{*}$. By employing this strategy, the detector who receives the recommendation to keep silent cannot distinguish whether this recommendation is caused by the truth-telling strategy of the principal in the good state or by his randomized strategy in the bad state. Therefore, she makes a belief about each of these two cases and takes an action that maximizes her expected continuation utility with respect to these beliefs. The probability $q_{n_p}^{*}$ is chosen as the maximum probability which makes obedience the best action for the detector. We referred to this region as the \emph{randomized region}

\textbf{Region 3:} In the third region which consists of times after the threshold $n_p^{*}$, the state of the Markov chain can be exactly derived from the principal's messages. In this time interval, a recommendation to keep silent means that the Markov chain is in the good state and a recommendation to declare the jump means that the state of the Markov chain has switched to the bad state. This region can be referred to as the \emph{full-information region}.

\begin{figure}[t]
\centering
\includegraphics[width=0.7 \textwidth, height=5cm]{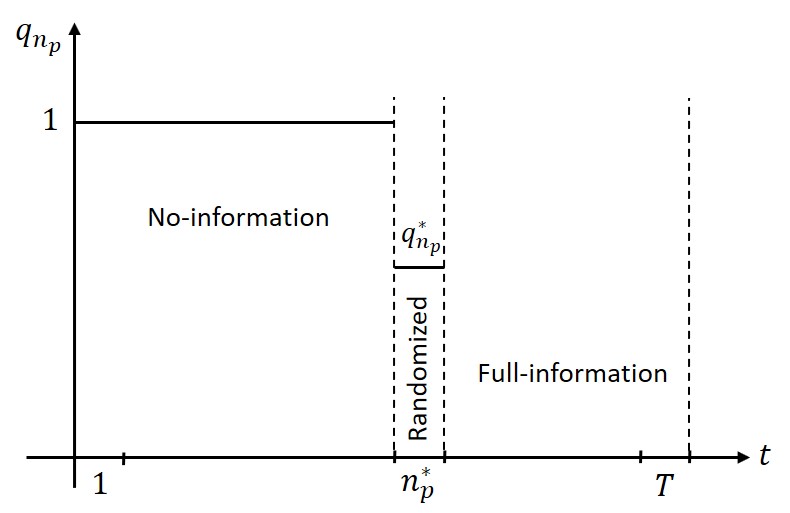}
\caption{The principal's strategy in the optimal mechanism}
\label{fig:optimal-mechanism}
\end{figure}

Based on the above arguments, we can depict the principal's optimal strategy as in Fig. \ref{fig:optimal-mechanism}. In this optimal mechanism, the principal does not give any information to the detector up to time $n_p^{*}-1$, but he promises that if the detector remains silent until that point, then he starts to provide him with ``almost accurate'' information. The information provided by the principal after time $n_p^{*}-1$ would contain some noise at time $n_p^{*}$, but will be precise and fully revealing of the state after that. 

The commitment of the principal to full disclosure of information after a certain time increases the detector's patience and gives her incentives to remain silent longer. To see this we compute the length of time the detector remains silent in two instances: (1) when the principal employs a no-information disclosure strategy for the whole horizon; (2) when the principal commits to full information disclosure some time in the future. In the first instance, the detector's expected cost if she declares the jump at $\tau=1, \ldots, T$ is 
\begin{align}\label{eq:cost-detector-no}
\begin{split}
&\mathbb{E}^{No} \{J^D(\tau,\theta)\}=\mathbb{P} (\theta >\tau )+c \sum_{t=1}^{\tau-1}{\mathbb{P} (\theta \leq t )}=\\
&\mu  (1-q)^{\tau-1}+c \sum_{t=1}^{\tau-1}{(1-\mu  (1-q)^{t-1})}=\\
&\mu (1-q)^{\tau-1}+c (\tau-1)- c  \mu  \frac{1-(1-q)^{\tau-1}}{q}.
\end{split}
\end{align}
If the detector keeps silent at the whole horizon, captured by $\tau=T+1$, her expected cost is
\begin{align}\label{eq:cost-detector-no-sil}
\mathbb{E}^{No} \{J^D(T+1,\theta)\}\hspace{-0.07cm}=\hspace{-0.07cm}c \sum_{t=1}^{T}{\mathbb{P} (\theta \leq t )}\hspace{-0.07cm}=\hspace{-0.07cm}cT\hspace{-0.07cm}-\hspace{-0.07cm}c \mu \frac{1\hspace{-0.07cm}-\hspace{-0.07cm}(1\hspace{-0.07cm}-\hspace{-0.07cm}q)^{T}}{q}.
\end{align}
Therefore, to minimize her expected cost, the detector with no additional information declares the jump at time 
\begin{align}\label{eq:cost-detector-no-sil2}
\tau^{No}=\argmin_{\tau=1:T+1}{\mathbb{E}^{No} \{J^D(\tau,\theta)\}}.\text{\footnotemark}
\end{align}
The case $\tau^{No}=T+1$ means that the belief $\mu$ the detector has in the good state of the Markov chain is high enough that the best action for her is to keep silent over the whole horizon. Using the optimal stopping time $\tau^{No}$, we can derive the principal's utility as follows:
\begin{align}\label{eq:utility-No}
\mathbb{E}^{No} \{U^P\}=\tau^{No}-1.
\end{align}
\footnotetext{For every fixed initial probability mass function, $\tau^{No}$ is a deterministic quantity.}

\begin{figure}
    \centering
    \begin{subfigure}[b]{0.28\textwidth}
        \includegraphics[height=0.77\textwidth]{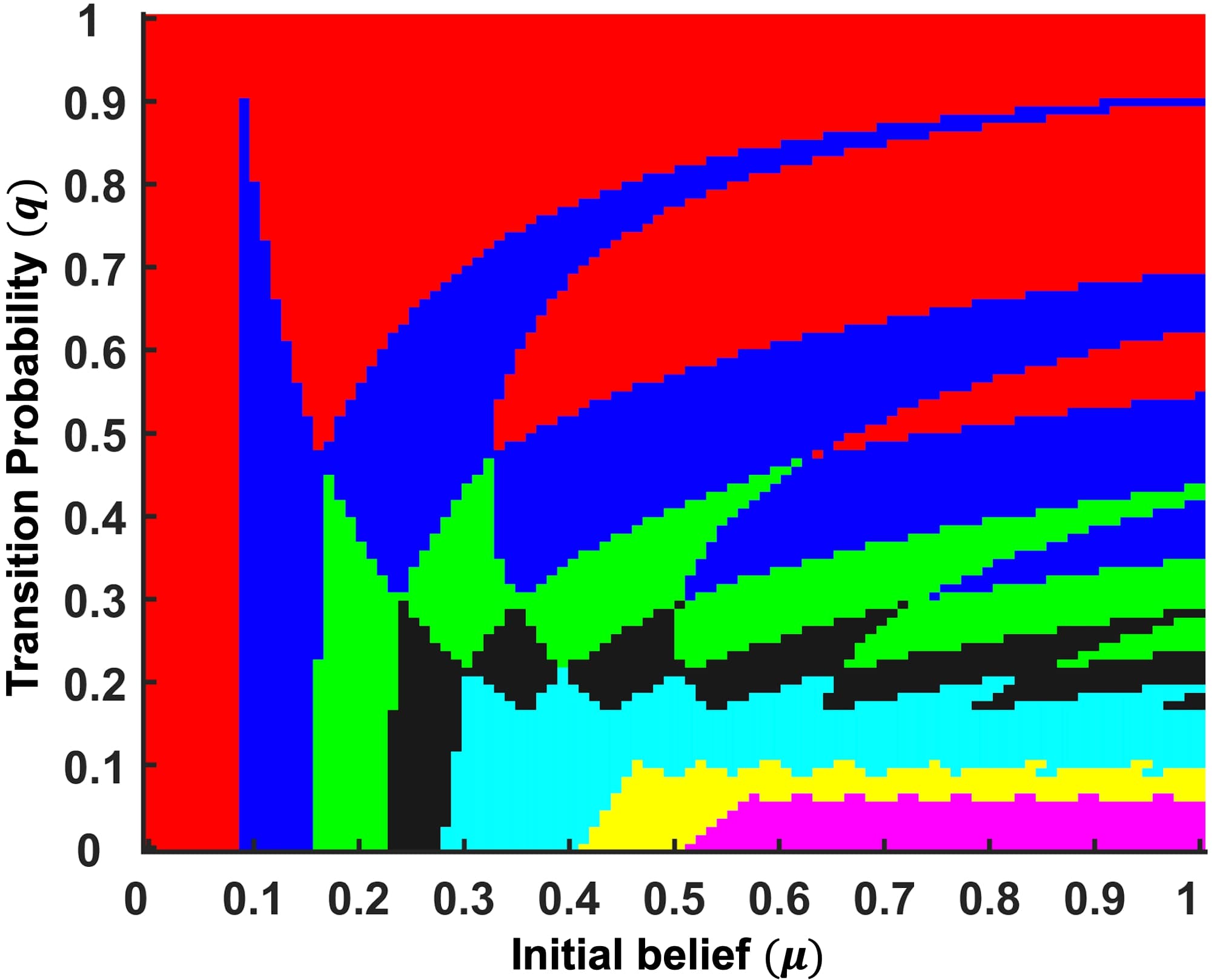}
        \caption{$c=0.1$}
        \label{fig:patience-c}
    \end{subfigure}
    ~ 
    \begin{subfigure}[b]{0.28\textwidth}
        \includegraphics[height=0.77\textwidth]{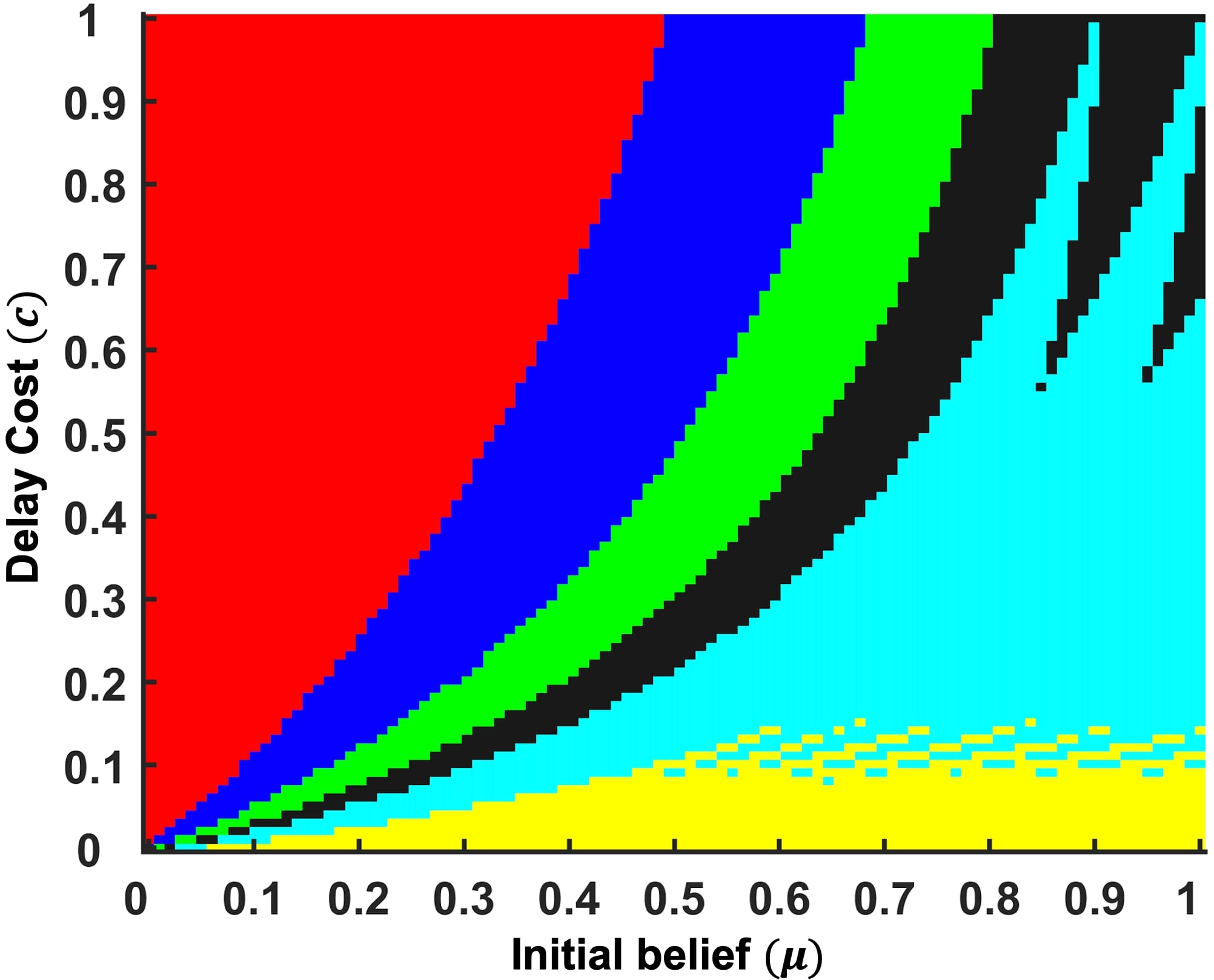}
        \caption{$q=0.1$}
        \label{fig:patience-q}
    \end{subfigure}
    ~ 
    \begin{subfigure}[b]{0.28\textwidth}
        \includegraphics[height=0.77\textwidth]{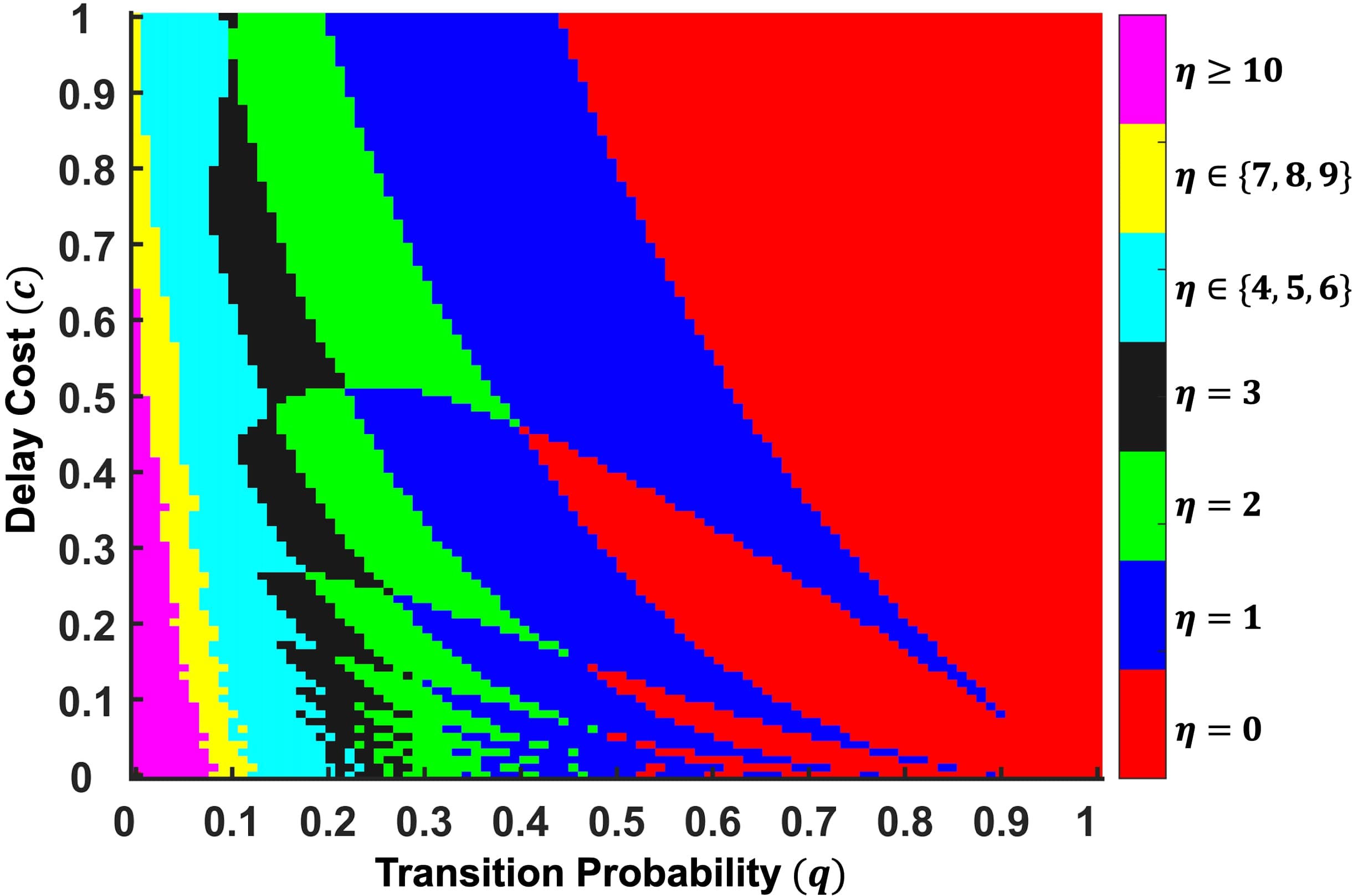}
        \caption{$\mu=0.9$}
        \label{fig:patience-mu}
    \end{subfigure}
    \caption{Color-coded map. Red: $\eta\hspace{-0.07cm}=\hspace{-0.07cm}0$, Blue: $\eta\hspace{-0.07cm}=\hspace{-0.07cm}1$, Green: $\eta\hspace{-0.07cm}=\hspace{-0.07cm}2$, Black: $\eta\hspace{-0.07cm}=\hspace{-0.07cm}3$, Cyan: $4 \hspace{-0.07cm} \leq \hspace{-0.07cm} \eta \hspace{-0.07cm} \leq \hspace{-0.07cm} 6$, Yellow: $7 \hspace{-0.07cm} \leq \hspace{-0.07cm} \eta \hspace{-0.07cm} \leq\hspace{-0.07cm}  9$, Magentic: $\eta \hspace{-0.07cm}\geq \hspace{-0.07cm} 10$.}
		\label{fig:patience}
\end{figure}

The arguments above show that the detector who receives no new information remains silent for $\tau^{No}-1$ numbers of time slots. In the optimal time-based-prioritized mechanism (instance 2), the principal's commitment to provide almost accurate information in the future incentivizes the detector to keep silent for $n_p^{*}-1$ periods of time without receiving any new information.  
The difference $\eta=n_p^{*}-\tau^{No}$, which we call the \emph{patience enhancement}, measures by how much the principal's commitment increases the detector's patience. In Fig. \ref{fig:patience}, we have plotted this measure for different values of the parameters $\mu \in [0,1]$, $q \in [0,1]$ and $c \in [0,1]$. In each sub-figure, we fix one of the parameters and partition the space of the two remaining parameters in terms of the patience enhancement $\eta$. Based on Fig. \ref{fig:patience-c}, the detector's patience increases by at least one time slot in $57.17\%$ of the cases, when $c=1$. For $c=1$, the patience enhancement is at least $4$, $7$ and $10$ time slots in $13.38\%$, $5.64\%$, and $2.94\%$ of the cases, respectively. The probabilities of having a patience enhancement above thresholds $1$, $4$, $7$, and $10$, are depicted in Table \ref{Table1} for each of sub-figures \ref{fig:patience-c}-\ref{fig:patience-mu}.
These results show that the principal's commitment to provide almost accurate information in the future can significantly enhance the detector's patience.


%
%
%

\begin{table}
\centering
\begin{tabular}{| c || c | c | c | c|}
 \hline
 & $\eta \geq 1$ & $\eta \geq 4$ & $\eta \geq 7$ & $\eta \geq 10$ \\ 
 \hline
 $c=0.1$ & $57.17\%$ & $13.38\%$ & $5.64\%$ & $2.94\%$\\  
 \hline
$q=0.1$ & $69.57\%$ & $31.12\%$ & $8.13\%$ & $0\%$\\  
 \hline
$\mu=0.9$ & $56.65\%$ & $12.83\%$ & $5.14\%$ & $2.39\%$\\  
 \hline
\end{tabular}
\caption{Probabilities of having a patience enhancement above certain thresholds}  \label{Table1}
\end{table}

\textbf{Feature 3}. In the optimal time-based prioritized mechanism derived by Algorithm 1, we choose the maximum value of $n_p$ such that there exists a $q_{n_p} \in [0,1]$ such that $\rho=(n_p,q_{n_p})$ satisfies all the obedience constraints for times before $n_p$. To do so, at each round, the algorithm considers a fix value of $n_p$ and computes the maximum value of $q_{n_p}$ that together with $n_p$ satisfies all the obedience constraints of times $t \leq n_p$. 
We can simplify Algorithm 1 by using the result of next theorem.

\begin{theorem}
	\label{T-binding}
	\normalfont
Suppose $n_p \geq \tau^{No}$. Let 
\begin{equation}\label{eq:binding1}
q_{n_p}^t=\frac{1}{\mathbb{P}(\theta \leq n_p)}(\frac{\mathbb{P}(\theta > t)}{c}-\sum_{l=t}^{n_p-1}{\mathbb{P}(\theta \leq l)}),
\end{equation}
denote the maximum value of $q_{n_p}$ that together with $n_p$ satisfies the obedience constraint of time $t$.
Then, we have
\begin{equation}\label{eq:binding2}
q_{n_p}^{\tau^{No}}=\min_{t \leq n_p}{q_{n_p}^t},
\end{equation}
where $\tau^{No}$ is the time at which the detector declares the jump, when the principal employs a no-information strategy.
\end{theorem}

Theorem \ref{T-binding} intuitively says that persuading the detector to keep silent at time $\tau^{No}$ is the most difficult challenge faced by the principal. This theorem suggests the following simplification for Algorithm 1. To find the optimal mechanism, we can run Algorithm 1 up to $n_p=\tau^{No}-1$. Then, if the terminating condition (Lines 3-4) has not been satisfied yet, we can replace Line 2 of the algorithm with
\begin{equation}\label{eq:binding3}
q_{n_p}=\frac{1}{\mathbb{P}(\theta \leq n_p)}(\frac{\mathbb{P}(\theta > \tau^{No})}{c}-\sum_{l=\tau^{No}}^{n_p-1}{\mathbb{P}(\theta \leq l)}),
\end{equation}
where $\tau^{No}$ is derived by (\ref{eq:cost-detector-no-sil2}). This simplification reduces the complexity of Algorithm 1, as the algorithm does not need to solve an optimization problem at each round, anymore. Instead, it can solve the optimization problem (\ref{eq:cost-detector-no-sil2}) once and use the result to find the maximum feasible value of $q_{n_p}$ at each round.

\textbf{Feature 4}. In Fig. \ref{fig:patience} and Table \ref{Table1}, we showed the superiority of our proposed mechanism compared to the no-information mechanism. In this part, we want to compare our mechanism with two other benchmark mechanisms, in terms of the expected utility they can provide for the principal. 

The first benchmark is the full information mechanism in which the principal reveals perfectly the Markov chain's state to the detector. The full information mechanism can be considered as a time-based prioritized mechanism with $n_p=1$, and $q_{n_p}=0$. Therefore, we can conclude from (\ref{eq:utility-time-pr}) and (\ref{eq:theta-dist}) that the expected utility the principal gets if he honestly shares his information with the detector is
\begin{align}\label{eq:utility-full}
\mathbb{E}^{Full} \{U^P\}\hspace{-0.07cm}=\hspace{-0.07cm}\sum_{t=1}^T{\mathbb{P}(\theta \hspace{-0.07cm} >\hspace{-0.07cm} t)}\hspace{-0.07cm}=\hspace{-0.07cm}\sum_{t=1}^T{\mu (1\hspace{-0.07cm}-\hspace{-0.07cm}q)^{t-1}}=\mu \frac{1\hspace{-0.07cm}-\hspace{-0.07cm}(1\hspace{-0.07cm}-\hspace{-0.07cm}q)^{T}}{q}.
\end{align}

The second benchmark we consider here, is the best static mechanism that can be employed by the principal. This comparison highlights the power of dynamic mechanisms compared to static ones. In a static mechanism, the set of messages $\mathcal{M}$ that the principal sends to the detector at each instant of time, as well as the distribution over $\mathcal{M}$ given the current state is time-independent. By the direct revelation principle, without loss of generality, we can focus on \emph{direct} static mechanisms in which the detector follows the principal's recommendations.

In a direct static mechanism, the principal recommends the detector to keep silent with probability $\rho_{s_t}$, where $s_t \in \{g,b\}$ is the current state of the Markov chain. By an argument similar to that in Lemma \ref{L-g-state}, we can show that in the optimal static mechanism, we have $\rho_{g}=1$. Moreover, we can show that the principal's expected utility is an increasing function of $\rho_{b}$.
Therefore, the problem of finding the best static mechanism is equivalent to finding the maximum value of $\rho_{b}$ such that the mechanism satisfies the obedience constraints. By some algebra, the detector's obedience constraint at each time $t$ can be derived as follows:
\begin{align}\label{eq:obed-static}
\frac{\mu (1-q)^{t-1}}{\mu (1-q)^{t-1}+(1-\mu)\rho_{b}^{t}+\sum_{\tau=1}^{t-1}{\mu (1-q)^{\tau-1}q \rho_b^{t-\tau}}} \geq \nonumber\\
c \sum_{l=t}^T{\frac{(1-\mu)\rho_{b}^{l}+\sum_{\tau=1}^{l-1}{\mu (1-q)^{\tau-1}q \rho_b^{l-\tau}}}{\mu (1-q)^{t-1}+(1-\mu)\rho_{b}^{t}+\sum_{\tau=1}^{t-1}{\mu (1-q)^{\tau-1}q \rho_b^{t-\tau}}}}
\end{align}
where the left-hand side is the average cost of declaring a jump and the right-hand side is the expected cost of keeping silent at time $t$, when the detector is recommended to keep silent. We denote the maximum value of $\rho_b \in [0,1]$ that satisfies constraint (\ref{eq:obed-static}) for each $t \leq T$, by $\hat{\rho}$. Therefore, an efficient static mechanism for the principal is a direct mechanism with $\rho_{g}=1$ and $\rho_{b}=\hat{\rho}$. This mechanism provides the principal with the following expected utility:
\begin{align}\label{eq:utility-static}
\mathbb{E}^{stat} \{U^P\}=\sum_{t=1}^T{[\mathbb{P}(\theta > t)+\sum_{\theta'=1}^t{\mathbb{P}(\theta =\theta') \hat{\rho}^{t-\theta'+1}}]}=\nonumber\\
\sum_{t=1}^T{[\mu \hspace{0.1cm} (1-q)^{t-1}+(1-\mu)\hat{\rho}^{t}+\sum_{\theta'=2}^t{\mu \hspace{0.1cm} (1-q)^{\theta'-2}\hspace{0.1cm} q \hspace{0.1cm} \hat{\rho}^{t-\theta'+1}}]}.
\end{align}

\begin{figure}[t]
\centering
\includegraphics[width=0.7 \textwidth, height=5cm]{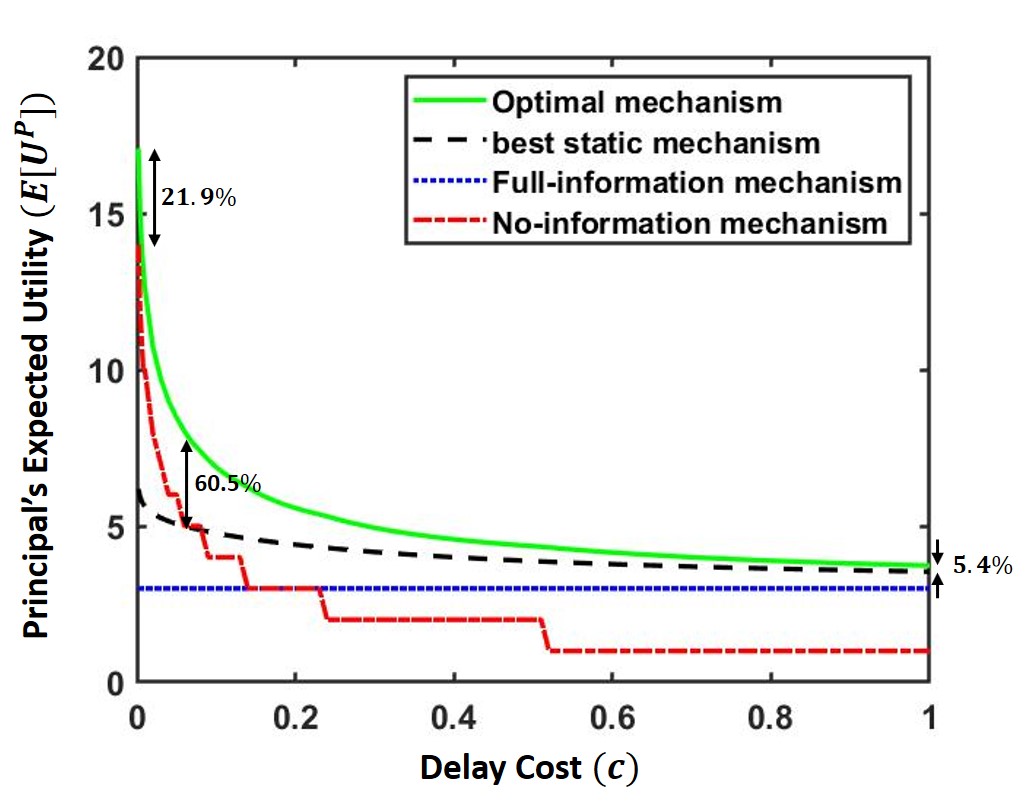}
\caption{Principal's expected utility vs. delay cost, for $\mu=0.9, q=0.3, T=50$.  }
\label{fig:compare-utility-c}
\end{figure}

In Fig. \ref{fig:compare-utility-c}, we have illustrated the principal's expected utility when he adopts the optimal dynamic, best static, full-information and no-information mechanisms, for different delay costs $c$, when $\mu=0.9$, $q=0.3$ and $T=50$. We observe that while the benchmark mechanisms outperform each other in different regions, the optimal mechanism proposed in this paper always outperforms all of them. In the comparison of different benchmarks, we can see that for low values of delay cost $c$, the no-information mechanism outperforms the full-information and best static mechanisms. This is simply because when the cost of delay is much lower than the false alarm fee, the detector with no additional information prefers to postpone detecting the jump so as to avoid detecting a false alarm. This behavior is in the principal's favor and provides him with a high expected utility. However, when the delay cost $c$ goes up, the detector's patience decreases and she prefers to detect the jump sooner. In this situation, the performance of the best static mechanism is superior to that of the two naive mechanisms described above (i.e., full-information and no-information mechanisms). It can be observed from Fig. \ref{fig:compare-utility-c} that the percentage of principal's utility enhancement in the optimal mechanism proposed in this paper compared to the best available benchmark starts from $21.9\%$ when $c=0$, goes up to $60.5\%$ in $c=0.06$ and then decreases gradually to $5.4\%$ when the delay cost reaches one.

\section{Conclusion} \label{sec:conclusion}

We studied a dynamic Bayesian persuasion problem whereby a strategic principal observes the evolution of a Markov chain and designs a recommendation policy that generates a recommendation to a strategic detector at each time. The goals of the principal and the detector are different, therefore, the long-term-optimizing detector does not have to obey the principal's recommendations unless she is convinced to do so. We presented a sequential recommendation policy that maximizes the principal's utility, and ensures the detector's obedience. We proved that the optimal policy is a threshold type, with two thresholds that can be explicitly computed. As time goes by, the optimal recommendation strategy first shifts from a no-information type to a randomized type and then switches to a full-information type.

\appendix
\section{Proof of Lemma \ref{L-message-history}}
We want to show that at each time $t$, part of the recommendation policy $\rho_{t}^{\theta,m_{1:t-1}}$ with $\mathcal{N}_{d}(m_{1:t-1})>1$ does not appear on either the obedience constraints or the principal's utility function. We first look at the obedience constraints (\ref{eq:obedience-constraint-k})-(\ref{eq:obedience-constraint-d}). Substituting (\ref{eq:cost-declare}), (\ref{eq:cost-silence}) and (\ref{eq:current-belief}) in (\ref{eq:obedience-constraint-k})-(\ref{eq:obedience-constraint-d}) shows that inequalities 
\begin{align}\label{eq:obedience-constraint-k-2}
c + \mathbb{E}^{\boldsymbol{\rho}}_{t+1:T} \{J^D(\tau,\theta) | m_{1:t}=(k)^t, a_t=k, a_{t+1:T}=m^{\boldsymbol{\rho}}_{t+1:T}\} \leq \nonumber\\
(1+c) \hspace{0.1cm} \frac{\mathbb{P}(\theta > t) \prod_{t'=1}^t{\rho_{t'}^{t'+1,(k)^{t'-1}}}}{\sum_{\theta'=1}^{T+1}{\mathbb{P} (\theta=\theta')\prod_{t'=1}^{t}{\rho_{t'}^{\min{(\theta',t'+1)},(k)^{t'-1}}}}}, \forall t \in \mathcal{T},
\end{align}
and 
\begin{align}\label{eq:obedience-constraint-d-2}
(1+c) \hspace{0.1cm} \frac{\mathbb{P}(\theta > t) \prod_{t'=1}^{t-1}{\rho_{t'}^{t'+1,(k)^{t'-1}}}(1-\rho_{t}^{t+1,(k)^{t-1}})}{\sum_{\theta'=1}^{T+1}{\mathbb{P} (\theta=\theta')\prod_{t'=1}^{t-1}{\rho_{t'}^{\min{(\theta',t'+1)},(k)^{t'-1}}}(1-\rho_{t'}^{\min{(\theta',t+1)},(k)^{t-1}})}} \leq \nonumber\\
c + \mathbb{E}^{\boldsymbol{\rho}}_{t+1:T} \{J^D(\tau,\theta) | m_{1:t}=((k)^{t-1},d), a_t=k, a_{t+1:T}=m^{\boldsymbol{\rho}}_{t+1:T}\}, \forall t \in \mathcal{T},
\end{align}
are equivalent to the obedience constraints (\ref{eq:obedience-constraint-k})-(\ref{eq:obedience-constraint-d}). The expected value of the detector's future costs from time $t+1$ onward when she obeys the recommendations is
\begin{align}\label{eq:future_cost-obey}
&\mathbb{E}^{\boldsymbol{\rho}}_{t+1:T} \{J^D(\tau,\theta) | m_{1:t}, a_t=k, a_{t+1:T}=m^{\boldsymbol{\rho}}_{t+1:T}\}= \nonumber\\
&\sum_{l=t+1}^T{\mathbb{P}(s_l=g, m^{\boldsymbol{\rho}}_{t+1:l-1}=(k)^{l-t-1},m^{\boldsymbol{\rho}}_l=d | m_{1:t})}+\nonumber\\
&\sum_{l=t+1}^T{c \hspace{0.1cm} \mathbb{P}(s_l=b, m^{\boldsymbol{\rho}}_{t+1:l}=(k)^{l-t} | m_{1:t})},
\end{align}
where the first term is the probability of occurring a false alarm and the second term is the expected cost of delay in the detection of the jump. Using Bayes' rule to calculate the probabilities appearing in (\ref{eq:future_cost-obey}) for $m_{1:t}=(k)^t$ and $m_{1:t}=((k)^{t-1},d)$, we obtain
\begin{align}\label{eq:future_cost-obey2}
&\mathbb{E}^{\boldsymbol{\rho}}_{t+1:T} \{J^D(\tau,\theta) | m_{1:t}=(k)^t, a_t=k, a_{t+1:T}=m^{\boldsymbol{\rho}}_{t+1:T}\}= \nonumber\\ 
&\sum_{l=t+1}^T{\frac{\mathbb{P}(\theta > l) \prod_{t'=1}^{l-1}{\rho_{t'}^{t'+1,(k)^{t'-1}}} (1-\rho_{l}^{l+1,(k)^{l-1}}) }{\sum_{\theta'=1}^{T+1}{\mathbb{P} (\theta=\theta')\prod_{t'=1}^{t}{\rho_{t'}^{\min{(\theta',t'+1)},(k)^{t'-1}}}}}}+\nonumber\\ 
&\sum_{l=t+1}^T{c \hspace{0.1cm} \frac{\sum_{\theta'=1}^l{\mathbb{P}(\theta=\theta') \prod_{t'=1}^{l}{\rho_{t'}^{\min{(\theta',t'+1)},(k)^{t'-1}}} }}{\sum_{\theta'=1}^{T+1}{\mathbb{P} (\theta=\theta')\prod_{t'=1}^{t}{\rho_{t'}^{\min{(\theta',t'+1)},(k)^{t'-1}}}}}},
\end{align}
and
\begin{equation}\label{eq:future_cost-obey3}
\mathbb{E}^{\boldsymbol{\rho}}_{t+1:T} \{J^D(\tau,\theta) | m_{1:t}=((k)^{t-1},d), a_t=k, a_{t+1:T}=m^{\boldsymbol{\rho}}_{t+1:T}\}=\frac{A}{B},
\end{equation}
where 
\begin{align}\label{eq:future_cost-obey3-AB}
\begin{split}
&A\hspace{-0.07cm}=\hspace{-0.07cm}\sum_{l=t\hspace{-0.04cm}+\hspace{-0.04cm}1}^T{\mathbb{P}(\theta \hspace{-0.05cm}>\hspace{-0.05cm}l) \hspace{-0.07cm} \prod_{t'=1}^{t-1}{\rho_{t'}^{t'\hspace{-0.05cm}+\hspace{-0.05cm}1,(k)^{t'\hspace{-0.05cm}-\hspace{-0.05cm}1}}} (1\hspace{-0.08cm}-\hspace{-0.08cm}\rho_{t}^{t\hspace{-0.05cm}+\hspace{-0.05cm}1,(k)^{t\hspace{-0.05cm}-\hspace{-0.05cm}1}}) \hspace{-0.07cm} \prod_{t'=t+1}^{l-1}{\rho_{t'}^{t'\hspace{-0.05cm}+\hspace{-0.05cm}1,(k)^{t'\hspace{-0.05cm}-\hspace{-0.05cm}1}_{-t}}}(1\hspace{-0.05cm}-\hspace{-0.05cm}\rho_{l}^{l\hspace{-0.03cm}+\hspace{-0.03cm}1,(k)^{l\hspace{-0.03cm}-\hspace{-0.03cm}1}_{-t}})+} \\
&\sum_{l=t\hspace{-0.04cm}+\hspace{-0.04cm}1}^T{\hspace{-0.07cm}\sum_{\theta'\hspace{-0.05cm}=\hspace{-0.05cm}1}^{l}{c \hspace{0.02cm} \mathbb{P}(\theta\hspace{-0.05cm}=\hspace{-0.05cm}\theta')\hspace{-0.1cm} \prod_{t'\hspace{-0.05cm}=\hspace{-0.05cm}1}^{t\hspace{-0.05cm}-\hspace{-0.05cm}1}{\rho_{t'}^{\min{\hspace{-0.05cm}(\theta'\hspace{-0.05cm},t'\hspace{-0.05cm}+\hspace{-0.05cm}1)},(k)^{t'\hspace{-0.05cm}-\hspace{-0.05cm}1}}}\hspace{-0.08cm} (1\hspace{-0.1cm}-\hspace{-0.1cm}\rho_{t}^{\min{\hspace{-0.05cm}(\theta'\hspace{-0.05cm},t+1)},(k)^{t\hspace{-0.05cm}-\hspace{-0.05cm}1}}) \hspace{-0.07cm} \hspace{-0.1cm} \prod_{t'\hspace{-0.03cm}=\hspace{-0.04cm}t\hspace{-0.05cm}+\hspace{-0.05cm}1}^{l}{\hspace{-0.05cm}\rho_{t'}^{\min{\hspace{-0.05cm}(\theta'\hspace{-0.05cm},t'\hspace{-0.05cm}+\hspace{-0.05cm}1)},(k)^{t'\hspace{-0.07cm}-\hspace{-0.07cm}1}_{-t}}}}}\hspace{-0.05cm},\\
&B=\sum_{\theta'=1}^{T+1}{\mathbb{P} (\theta=\theta')\prod_{t'=1}^{t-1}{\rho_{t'}^{\min{(\theta',t'+1)},(k)^{t'-1}}}(1-\rho_{t'}^{\min{(\theta',t+1)},(k)^{t-1}})},
\end{split}
\end{align}
and $(k)^{t'}_{-t}$ is a vector of length $t'$ where all the components expect the $t$-th one equal $k$, and the $t$-th component is $d$. Substituting (\ref{eq:future_cost-obey2}) and (\ref{eq:future_cost-obey3}) in (\ref{eq:obedience-constraint-k-2})-(\ref{eq:obedience-constraint-d-2}), we can see that the terms of the recommendation policy that appear in the obedience constraints have at most one $d$ in their message history. Furthermore, (\ref{eq:principal-utility}) shows that the principal's utility function depends only on the recommendation probabilities $\rho_{t}^{\theta_t,m_{1:t-1}}$ where the message history is $m_{1:t-1}=(k)^{t-1}$. Hence the proof of Lemma \ref{L-message-history} is complete.

\section{Proof of Lemma \ref{L-state-history}}
For each time $t$, $\theta=1,\ldots,t$ means that the jump has occurred and the state of the Markov chain becomes bad, i.e. $s_t=b$; furthermore, $\theta=t+1$ captures the fact that the jump has not occurred yet and the state is still good, i.e. $s_t=g$. Therefore, to prove Lemma \ref{L-state-history}, we need to show that it is without loss of optimality if the principal restricts attention to recommendation policies with the same $\rho_{t}^{\theta,m_{1:t-1}}$ for all $\theta=1, \ldots, t$. We claim that in this class of mechanisms, the recommendations do not depend on the exact time of the jump, but only on whether it has occurred. 
To prove the claim, we show that for each $t \in \mathcal{T}$ and each $m_{1:t-1}$ with $\mathcal{N}_{d}(m_{1:t-1}) \leq 1$, the parameters $\rho_{t}^{\theta,m_{1:t-1}}$ with $\theta=1, \ldots, t$ always appear in the obedience constraints and the principal's utility function as parts of a constant linear combination. Therefore, as long as the value of this linear combination is fixed, the values of the individual parameters can be customized. As a result, for each optimal mechanism, we can construct another optimal mechanism in which for each $t$ and each $m_{1:t-1}$, the values of all parameters $\rho_{t}^{\theta,m_{1:t-1}}$, with $\theta=1,\ldots, t$, are the same.

We proceed by backward induction on time $t$.

\textbf{Basis of induction.} Let $t=T$. The recommendation policy of the final time $T$ appears in the obedience constraints of time $T$ and of times $t'<T$. We start our investigation by studying the obedience constraints of time $T$. According to (\ref{eq:obedience-constraint-k})-(\ref{eq:cost-silence}) and due to the fact that there is no future after time $T$, the obedience constraints of time $T$ are as follows:
\begin{align}\label{eq:obedience-constraint-k-T}
c \hspace{0.1cm} (1-\mathbb{P} (s_T=g | m_{1:T}=(k)^T)) \leq \mathbb{P} (s_T=g | m_{1:T}=(k)^T),
\end{align}
and
\begin{align}\label{eq:obedience-constraint-d-T}
\mathbb{P} (s_T=g | m_{1:T}=((k)^{t-1},d)) \leq c \hspace{0.1cm} (1-\mathbb{P} (s_T=g | m_{1:T}=((k)^{t-1},d))).
\end{align}
By substituting (\ref{eq:current-belief}) in (\ref{eq:obedience-constraint-k-T}) and (\ref{eq:obedience-constraint-d-T}) and canceling out the denominators, we can rewrite the obedience constraints of time $T$ as follows:
\begin{align}\label{eq:obedience-constraint-k-T3}
c \left[\sum_{\theta'=1}^T{\mathbb{P}(\theta=\theta') \prod_{t'=1}^{\theta'-1}{\rho_{t'}^{t'+1,(k)^{t'-1}}} \prod_{t'=\theta'}^T{\rho_{t'}^{\theta',(k)^{t'-1}}}}\right] \leq \mathbb{P}(s_{T}=g) \prod_{t'=1}^T{\rho_{t'}^{t'+1,(k)^{t'-1}}},
\end{align}
and
\begin{align}\label{eq:obedience-constraint-d-T3}
&\mathbb{P}(s_{T}=g) \prod_{t'=1}^{T-1}{\rho_{t'}^{t'+1,(k)^{t'-1}}} (1-\rho_{T}^{T+1,(k)^{T-1}}) \leq \nonumber\\
&c \hspace{0.02cm} \left[\sum_{\theta'=1}^T{\mathbb{P}(\theta=\theta') \prod_{t'=1}^{\theta'-1}{\rho_{t'}^{t'+1,(k)^{t'-1}}} \prod_{t'=\theta'}^{T-1}{\rho_{t'}^{\theta',(k)^{t'-1}}}(1-\rho_{T}^{\theta',(k)^{T-1}})}\right].
\end{align}
It can be seen that in both constraints, the parameters $\rho_{T}^{\theta,(k)^{T-1}}$ for $\theta=1, \ldots, T$ only appeared in the following linear combination
\begin{align}\label{eq:linear-T}
\sum_{\theta'=1}^T{ \rho_{T}^{\theta',(k)^{T-1}} \hspace{0.1cm} \mathbb{P}(\theta=\theta') \prod_{t'=1}^{\theta'-1}{\rho_{t'}^{t'+1,(k)^{t'-1}}} \prod_{t'=\theta'}^{T-1}{\rho_{t'}^{\theta',(k)^{t'-1}}}},
\end{align}
which we denote by $L(\rho_{T}^{1,(k)^{T-1}},\ldots, \rho_{T}^{T,(k)^{T-1}})$.

Now we investigate how the parameters $\rho_{T}^{\theta,(k)^{T-1}}$ for $\theta=1, \ldots, T$ appear in the obedience constraints of time $t<T$. According to (\ref{eq:obedience-constraint-k})-(\ref{eq:obedience-constraint-d}), the obedience constraints of time $t<T$ are as follows:
\begin{align}\label{eq:obedience-constraint-k-t}
&c \hspace{0.1cm} \mathbb{P} (s_t=b | m_{1:t}=(k)^t) +  \nonumber\\ 
&\sum_{l=t+1}^T{\mathbb{P}(s_l=g, m_{t+1:l-1}=(k)^{l-t-1},m_l=d | m_{1:t}=(k)^t)}+ \nonumber\\ 
&\sum_{l=t+1}^T{c \hspace{0.1cm} \mathbb{P}(s_l=b, m_{t+1:l}=(k)^{l-t} | m_{1:t}=(k)^t)} \leq \mathbb{P} (s_t=g | m_{1:t}=(k)^t),
\end{align}
and
\begin{align}\label{eq:obedience-constraint-d-t}
&\mathbb{P} (s_t=g | m_{1:t}=((k)^{t-1},d)) \leq c \hspace{0.05cm} \mathbb{P} (s_t=b | m_{1:t}=((k)^{t-1},d)) + \nonumber\\
&\sum_{l=t+1}^T{\mathbb{P}(s_l=g, m_{t+1:l-1}=(k)^{l-t-1},m_l=d | m_{1:t}=((k)^{t-1},d))}+ \nonumber\\ 
&\sum_{l=t+1}^T{c \hspace{0.05cm} \mathbb{P}(s_l=b, m_{t+1:l}=(k)^{l-t} | m_{1:t}=((k)^{t-1},d))}.
\end{align}
We note that the distribution of $s_t$ given the messages received up to time $t$ does not depend on the future policies $\rho_{t+1:T}$. For each $l<T$, the joint distribution of the state $s_l$ and the messages received between $t+1$ and $l$, does not either depend on the recommendation policy of time $T$. Therefore, the only terms in (\ref{eq:obedience-constraint-k-t}) and (\ref{eq:obedience-constraint-d-t}) that depend on the recommendation policy of time $T$ are the terms appear in the summations with index $l=T$. The joint probability that the state $s_T$ is good and a certain message sequence $m_{t+1:T}$ happens between $t+1$ and $T$ depends on the recommendation policy the principal adopts at time $T$ when the jump has not occurred, that is $\theta=T+1$. Therefore, the only terms in the obedience constraints (\ref{eq:obedience-constraint-k-t}) and (\ref{eq:obedience-constraint-d-t}) that depend on parameters $\rho_{T}^{\theta,m_{1:T-1}}$, $\theta=1, \ldots, T$, are 
\begin{align}\label{eq:pr-L2-1}
&\mathbb{P}(s_T=b, m_{t+1:T}=(k)^{T-t} | m_{1:t}=(k)^t)=\nonumber\\
&\frac{1}{\mathbb{P} (m_{1:t}=(k)^t)} (\sum_{\theta'=1}^T{\mathbb{P}(\theta=\theta') \prod_{t'=1}^{\theta'-1}{\rho_{t'}^{t'+1,(k)^{t'-1}}} \prod_{t'=\theta'}^T{\rho_{t'}^{\theta',(k)^{t'-1}}}})
\end{align}
and
\begin{align}\label{eq:pr-L2-2}
&\mathbb{P}(s_T=b, m_{t+1:T}=(k)^{T-t} | m_{1:t}=((k)^{t-1},d))=\frac{1}{\mathbb{P} (m_{1:t}=((k)^{t-1},d))}\nonumber\\
&(\sum_{\theta'=1}^{t-1}{\mathbb{P}(\theta\hspace{-0.05cm}=\hspace{-0.05cm}\theta') \hspace{-0.05cm} \prod_{t'=1}^{\theta'\hspace{-0.05cm}-\hspace{-0.05cm}1}{\rho_{t'}^{t'\hspace{-0.05cm}+\hspace{-0.05cm}1,(k)^{t'\hspace{-0.05cm}-\hspace{-0.05cm}1}}} \hspace{-0.05cm} \prod_{t'=\theta'}^{t-1}{\rho_{t'}^{\theta',(k)^{t'-1}}} (1-\rho_{t}^{\theta',(k)^{t-1}}) \prod_{t'=t+1}^{T}{\rho_{t'}^{\theta',(k)^{t'-1}_{-t}}}}+\nonumber\\
&\mathbb{P}(\theta=t) \prod_{t'=1}^{t-1}{\rho_{t'}^{t'+1,(k)^{t'-1}}} (1-\rho_{t}^{t,(k)^{t-1}}) \prod_{t'=t+1}^{T}{\rho_{t'}^{t,(k)^{t'-1}_{-t}}}+\nonumber\\
&\sum_{\theta'=t+1}^{T}{\mathbb{P}(\theta\hspace{-0.05cm}=\hspace{-0.05cm}\theta') \hspace{-0.05cm} \prod_{t'=1}^{t\hspace{-0.05cm}-\hspace{-0.05cm}1}{\rho_{t'}^{t'+1,(k)^{t'\hspace{-0.05cm}-\hspace{-0.05cm}1}}} (1\hspace{-0.05cm}-\hspace{-0.05cm}\rho_{t}^{t+1,(k)^{t\hspace{-0.05cm}-\hspace{-0.05cm}1}}) \hspace{-0.05cm} \prod_{t'=t+1}^{\theta'\hspace{-0.05cm}-\hspace{-0.05cm}1}{\rho_{t'}^{t'\hspace{-0.05cm}+\hspace{-0.05cm}1,(k)^{t'\hspace{-0.05cm}-\hspace{-0.05cm}1}_{-t}}} \hspace{-0.05cm} \prod_{t'=\theta'}^{T}{\rho_{t'}^{\theta',(k)^{t'\hspace{-0.05cm}-\hspace{-0.05cm}1}_{-t}}}}),
\end{align}
respectively. Equation (\ref{eq:pr-L2-1}) is a multiple of $L(\rho_{T}^{1,(k)^{T-1}},\ldots, \rho_{T}^{T,(k)^{T-1}})$. Therefore, we can change the individual values of $\rho_{T}^{\theta,(k)^{T-1}}$, $\theta=1, \ldots, T$, without violating the obedience constraint (\ref{eq:obedience-constraint-k-t}), as long as the value of the function $L$ remains constant. Equation (\ref{eq:pr-L2-2}) includes parts of recommendation policy of time $T$ which correspond to message history $(k)^{T-1}_{-t}$. These parameters do not appear in any other constraint or in principal's utility function. Therefore, as long as the probability $\mathbb{P}(s_T=b, m_{t+1:T}=(k)^{T-t} | m_{1:t}=((k)^{t-1},d))$, given by (\ref{eq:pr-L2-2}), remains constant, changing the values of parameters $\rho_{T}^{\theta,(k)^{T-1}_{-t}}$, $\theta=1, \ldots, T$ does not violate the obedience property of the mechanism. 

The last point we should check is the effect of parameters $\rho_{T}^{\theta,m_{1:T-1}}$, $\theta=1, \ldots, T$, on the principal's utility. According to (\ref{eq:principal-utility}), the principal's utility is
\begin{align}\label{eq:principal-utility2}
&\mathbb{E}^{1:T} \{U^P(\tau)| \boldsymbol{\rho}\}=\nonumber\\
&\sum_{\theta'=1}^{T+1}{\mathbb{P}(\theta\hspace{-0.05cm}=\hspace{-0.05cm}\theta') \hspace{-0.05cm} \left[\sum_{t=1}^{\theta'-1}{\prod_{t'=1}^{t}{\rho_{t'}^{t'+1,(k)^{t'-1}}}}+\sum_{t=\theta'}^T{(\prod_{t'=1}^{\theta'-1}{\rho_{t'}^{t'+1,(k)^{t'-1}}\prod_{t'=\theta'}^{t}{\rho_{t'}^{\theta',(k)^{t'-1}}}})}\right]}=\nonumber\\
&\sum_{\theta'=1}^{T+1}{\mathbb{P}(\theta\hspace{-0.05cm}=\hspace{-0.05cm}\theta') \hspace{-0.05cm} \left[\sum_{t=1}^{\theta'-1}{\prod_{t'=1}^{t}{\rho_{t'}^{t'+1,(k)^{t'-1}}}}+\sum_{t=\theta'}^{T-1}{(\prod_{t'=1}^{\theta'-1}{\rho_{t'}^{t'+1,(k)^{t'-1}}\prod_{t'=\theta'}^{t}{\rho_{t'}^{\theta',(k)^{t'-1}}}})}\right]}+\nonumber\\
&\sum_{\theta'=1}^T{ \rho_{T}^{\theta',(k)^{T-1}} \hspace{0.1cm} \mathbb{P}(\theta=\theta') \prod_{t'=1}^{\theta'-1}{\rho_{t'}^{t'+1,(k)^{t'-1}}} \prod_{t'=\theta'}^{T-1}{\rho_{t'}^{\theta',(k)^{t'-1}}}}
\end{align}
It can be seen that the parameters $\rho_{T}^{\theta,(k)^{T-1}}$ for $\theta=1, \ldots, T$ do not appear in the first additive term of (\ref{eq:principal-utility2}). These parameters only appeared in the second component of (\ref{eq:principal-utility2}) as parts of the linear function $L(\rho_{T}^{1,(k)^{T-1}},\ldots, \rho_{T}^{T,(k)^{T-1}})$. Therefore, for each optimal mechanism we can construct another optimal mechanism where for each message sequence $m_{1:T-1}$, the values of parameters $\rho_{T}^{\theta,m_{1:t-1}}$, $\theta=1, \ldots, T$, are the same. For each $m_{1:T-1}$, we denote this common value by $\rho_{T}^{b,m_{1:T-1}}$, $\theta=1, \ldots, T$.

\textbf{Induction step.}  Suppose that the statement of the lemma is true for times after $t$. Now, we want to prove that it is also true for time $t$. To prove this fact, we employ an approach similar to the one used in proving the basis of induction. We show that for each $m_{1:t-1}$, parameters $\rho_{t}^{\theta,m_{1:t-1}}$ with $\theta=1, \ldots, t$ appear in all the obedience constraints and the principal's utility function as parts of a constant linear combination. To do so, we first study the obedience constraints of time $t$. The obedience constraints of time $t$ are as in (\ref{eq:obedience-constraint-k-t})-(\ref{eq:obedience-constraint-d-t}). By some basic algebra, one can show that the parameters $\rho_{t}^{\theta,m_{1:t-1}}$ with $\theta=1, \ldots, t$, appear in (\ref{eq:obedience-constraint-k-t}) as the following linear combination
\begin{align}\label{eq:obedience-constraint-k-t1}
\sum_{\theta'=1}^t{\rho_{t}^{\theta',(k)^{t-1}} \mathbb{P}(\theta=\theta') \prod_{t'=1}^{\theta'-1}{\rho_{t'}^{t'+1,(k)^{t'-1}}} \prod_{t'=\theta'}^{t-1}{\rho_{t'}^{\theta',(k)^{t'-1}}} (\sum_{l=t}^T{\prod_{t'=t+1}^{l}{\rho_{t'}^{\theta',(k)^{t'-1}}}})}.
\end{align}
By the induction hypothesis, the probabilities of recommending silence at times greater than $t$ do not depend on the exact time of the jump. Therefore, for each $t'\geq t+1$ we can replace $\rho_{t'}^{\theta',(k)^{t'-1}}$ by $\rho_{t'}^{b,(k)^{t'-1}}$. This replacement makes the last multiplicative term of (\ref{eq:obedience-constraint-k-t1}) independent of $\theta'$; hence we can neglect it. Therefore, the simplified form of the linear combination in which parameters $\rho_{t}^{\theta,(k)^{t-1}}$ with $\theta=1, \ldots, t$ appear is
\begin{align}\label{eq:linear-t}
\sum_{\theta'=1}^t{\rho_{t}^{\theta',(k)^{t-1}} \mathbb{P}(\theta=\theta') \prod_{t'=1}^{\theta'-1}{\rho_{t'}^{t'+1,(k)^{t'-1}}} \prod_{t'=\theta'}^{t-1}{\rho_{t'}^{\theta',(k)^{t'-1}}}}.
\end{align}
We denote this linear combination by $L(\rho_{t}^{1,(k)^{t-1}},\ldots, \rho_{t}^{t,(k)^{t-1}})$.
Now we investigate the second obedience constraint (\ref{eq:obedience-constraint-d-t}) of time $t$. We can show that the parameters $\rho_{t}^{\theta,m_{1:t-1}}$ with $\theta=1, \ldots, t$, appear in (\ref{eq:obedience-constraint-d-t}) as the following linear combination
\begin{align}\label{eq:obedience-constraint-d-t1}
\sum_{\theta'=1}^t{\rho_{t}^{\theta',(k)^{t-1}} \mathbb{P}(\theta=\theta') \prod_{t'=1}^{\theta'-1}{\rho_{t'}^{t'+1,(k)^{t'-1}}} \prod_{t'=\theta'}^{t-1}{\rho_{t'}^{\theta',(k)^{t'-1}}} (\sum_{l=t}^T{\prod_{t'=t+1}^{l}{\rho_{t'}^{\theta',(k)^{t'-1}_{-t}}}})}.
\end{align}
By the induction hypothesis, we have $\rho_{t'}^{\theta',(k)^{t'-1}_{-t}}=\rho_{t'}^{b,(k)^{t'-1}_{-t}}$, for $t' \geq t+1$. Using this equality, the last multiplicative term of (\ref{eq:obedience-constraint-d-t1}) becomes independent of $\theta'$; hence can be neglected. This makes the linear combination of (\ref{eq:obedience-constraint-d-t1}) exactly the same as $L(\rho_{t}^{1,(k)^{t-1}},\ldots, \rho_{t}^{t,(k)^{t-1}})$. Therefore, the parameters $\rho_{t}^{\theta,(k)^{t-1}}$ with $\theta=1, \ldots, t$ appear in the obedience constraints of time $t$ as parts of one single linear combination.

There are two other sets of obedience constraints that we should study: obedience constraints at times before $t$ and obedience constraints at times after $t$. Following the same steps as above, we can show that the parameters $\rho_{t}^{\theta,(k)^{t-1}}$, with $\theta=1, \ldots, t$, appear in all obedience constraints as well as the principal's utility function as parts of the linear function $L(\rho_{t}^{1,(k)^{t-1}},\ldots, \rho_{t}^{t,(k)^{t-1}})$. Therefore, as long as the value of this linear combination remains constant, changing the values of the individual parameters $\rho_{t}^{\theta,(k)^{t-1}}$, $\theta=1, \ldots, T$ does not violate the obedience constraints or change the principal's benefit. Other parameters of the recommendation policy of time $t$ that appear in the obedience constraints are $\rho_{t}^{\theta,(k)^{t-1}_{-t^{'}}}$, where $\theta=1, \ldots, t$ and $t'=1, \ldots, t-1$. For each $t'=1, \ldots, t-1$, the parameters $\rho_{t}^{\theta,(k)^{t-1}_{-t'}}$, with $\theta=1, \ldots, t$, appear as a linear combination only in the obedience constraint of time $t'$ when $m_{t'}=d$. Therefore, as long as the value of this linear combination remains constant, changing the values of parameters $\rho_{T}^{\theta,(k)^{t-1}_{-t'}}$, $\theta=1, \ldots, t$ does not make any difference. Combining all the arguments above we conclude that, for each optimal mechanism we can construct another optimal mechanism such that for each message sequence $m_{1:t-1}$, the values of parameters $\rho_{t}^{\theta,m_{1:t-1}}$, $\theta=1, \ldots, t$, are the same. For each $m_{1:t-1}$, we denote this common value by $\rho_{t}^{b,m_{1:t-1}}$, $\theta=1, \ldots, t$.

\textbf{Conclusion.} By the principle of induction, the statement is true for all $t=1, \ldots, T$. This completes the proof of lemma \ref{L-state-history}.

\section{Proof of Lemma \ref{L-g-state}}
We prove this Lemma by showing that we can increase the utility of principal in an obedient mechanism by setting $\rho_{t+1}^{g,m_{1:t}}=1$, for any $t \in \mathcal{T}$, and this change does not violate any of the obedience constraints. Having $\rho_{t+1}^{g,m_{1:t}}=1$ guarantees that the recommendations to declare the jump always will be obeyed by the detector. Therefore, in the following we focus on the recommendations to keep silent and show that by this change, the detector will be more willing to obey this kind of recommendations.

We proved in Lemma \ref{L-state-history} that, without loss of optimality, at any time $t \in \mathcal{T}$ the principal can restrict attention to recommendation policies that depend only on the message profile $m_{1:t-1}$ and the current state $s_t$ of the Markov chain (not on the state evolution). For any fixed recommendation policy $\boldsymbol{\rho}$ of the form suggested by Lemma \ref{L-state-history}, the detector is faced with a quickest detection problem \cite{Shiryaev}. However, there is a subtle difference between the standard quickest detection problem \cite{Shiryaev} and the one arising in our setup. In the standard quickest detection problem the detector's decision at any time $t$ depends only on her posterior belief $\hat{\pi}_t=\mathbb{P}(s_t=g|m_{1:t})$ about the state of the Markov chain at $t$. In the quickest detection problem arising in our setup, the detector's decision at any time $t$ depends on the received messages $m_{1:t}$, the principal's recommendation policy $\boldsymbol{\rho}$, and her belief 
\begin{align}\label{eq:belief}
\pi_t=\mathbb{P}(s_t=g|m_{1:t},\boldsymbol{\rho}).
\end{align}
This difference is due to the fact that at any time $t$ the principal's fixed recommendation policy $\boldsymbol{\rho}_{t+1}$ depends on $m_{1:t}$, therefore, the detector needs keep $m_{1:t}$ in order to determine the statistics of $m_{t+1}$ so as to update her information at time $t+1$. Consequently, in our setup, the detector's information state at any time $t$ is $(\pi_t,m_{1:t},\boldsymbol{\rho})$. The detector's optimal strategy $\gamma^*=(\gamma_1^*,\ldots, \gamma_T^*)$ is determined by the dynamic program
\begin{align}\label{eq:value-function-T}
W_T(\pi_T,m_{1:T},\boldsymbol{\rho})=\min{\left[\pi_T,c \hspace{0.05cm} (1-\pi_T)\right]},
\end{align}
\begin{align}\label{eq:value-function1}
&W_t(\pi_t,m_{1:t},\boldsymbol{\rho})=\nonumber\\
&\min{\left[\pi_t,c \hspace{0.05cm} (1\hspace{-0.05cm}-\hspace{-0.05cm}\pi_t)+\mathbb{E}\{W_{t+1}(\pi_{t+1},m_{1:t},m_{t+1},\boldsymbol{\rho}) | \pi_t,m_{1:t},\boldsymbol{\rho}\} \right]}, t\hspace{-0.05cm}=\hspace{-0.05cm}1, \ldots, T\hspace{-0.05cm}-\hspace{-0.05cm}1,
\end{align}
where
\begin{align}\label{eq:value-function2}
\pi_{t+1}=T_t(\pi_t,m_{1:t},m_{t+1},\boldsymbol{\rho}),
\end{align}
$T_t(.)$ is determined by Bayes' rule, and $m_{t+1}$ is a random variable that takes values in the set $\left\{d,k\right\}$; the statistics of $m_{t+1}$ are determined by $m_{1:t}$, the state $s_{t+1}$ of the Markov chain at time $t+1$, and the recommendation policy $\boldsymbol{\rho}$. The first term on the right hand side (RHS) of (\ref{eq:value-function-T}) and (\ref{eq:value-function1}) represents the detector's expected cost due to her decision to stop at $t$ and declare that jump has occurred $(a_t=1)$; the second term on the RHS of (\ref{eq:value-function-T}) and (\ref{eq:value-function1}) represents the detector's expected cost due to her decision to wait/remain silent at time $t$ $(a_t=0)$. The second term on the RHS of (\ref{eq:value-function1}) is equal to 
\begin{align}\label{eq:value-function3}
c \hspace{0.05cm} (1-\pi_t)+\mathbb{P} (m_{t+1}=k |\pi_t,m_{1:t},\boldsymbol{\rho}) W_{t+1}(T_t(\pi_t,m_{1:t},k,\boldsymbol{\rho}),m_{1:t},k,\boldsymbol{\rho})+ \nonumber\\
\mathbb{P} (m_{t+1}=d |\pi_t,m_{1:t},\boldsymbol{\rho}) W_{t+1}(T_t(\pi_t,m_{1:t},d,\boldsymbol{\rho}),m_{1:t},d,\boldsymbol{\rho}).
\end{align}
Furthermore, 
\begin{align}\label{eq:value-function4}
\mathbb{P} (m_{t+1}=k |\pi_t,m_{1:t},\boldsymbol{\rho})=\pi_t(1-q) \rho_{t+1}^{g,m_{1:t}}+(1-\pi_t(1-q)) \rho_{t+1}^{b,m_{1:t}}.
\end{align}
Using (\ref{eq:value-function1}) and (\ref{eq:value-function4}) and Bayes' rule to write explicitly $T_t(\pi_t,m_{1:t},k,\boldsymbol{\rho})$ and $T_t(\pi_t,m_{1:t},d,\boldsymbol{\rho})$, we obtain
\begin{align}\label{eq:value-function5}
\begin{split}
&W_t(\pi_t,m_{1:t},\boldsymbol{\rho})=
\min{\left[\pi_t,c \hspace{0.05cm} (1\hspace{-0.05cm}-\hspace{-0.05cm}\pi_t)+ (\pi_t(1\hspace{-0.05cm}-\hspace{-0.05cm}q) \rho_{t+1}^{g,m_{1:t}}+(1\hspace{-0.05cm}-\hspace{-0.05cm}\pi_t(1\hspace{-0.05cm}-\hspace{-0.05cm}q)) \rho_{t+1}^{b,m_{1:t}})\right.} \\
&{\left. W_{t+1}(\frac{\pi_t(1-q) \rho_{t+1}^{g,m_{1:t}}}{\pi_t(1-q) \rho_{t+1}^{g,m_{1:t}}+(1-\pi_t(1-q)) \rho_{t+1}^{b,m_{1:t}}},m_{1:t},k,\boldsymbol{\rho})+ \right.} \\
&{\left.  (\pi_t(1\hspace{-0.1cm}-\hspace{-0.1cm}q) (1\hspace{-0.1cm}-\hspace{-0.1cm}\rho_{t+1}^{g,m_{1:t}})\hspace{-0.1cm}+\hspace{-0.1cm}(1\hspace{-0.1cm}-\hspace{-0.1cm}\pi_t(1\hspace{-0.1cm}-\hspace{-0.1cm}q)) (1\hspace{-0.1cm}-\hspace{-0.1cm}\rho_{t+1}^{b,m_{1:t}})) \right.} \\
&{\left. W_{t+1}(\frac{\pi_t(1-q) (1\hspace{-0.1cm}-\hspace{-0.1cm}\rho_{t+1}^{g,m_{1:t}})}{\pi_t(1\hspace{-0.1cm}-\hspace{-0.1cm}q) (1\hspace{-0.1cm}-\hspace{-0.1cm}\rho_{t+1}^{g,m_{1:t}})\hspace{-0.1cm}+\hspace{-0.1cm}(1\hspace{-0.1cm}-\hspace{-0.1cm}\pi_t(1\hspace{-0.1cm}-\hspace{-0.1cm}q)) (1\hspace{-0.1cm}-\hspace{-0.1cm}\rho_{t+1}^{b,m_{1:t}})},m_{1:t},d,\boldsymbol{\rho})\right]}.
\end{split}
\end{align}
We use (\ref{eq:value-function-T}) and (\ref{eq:value-function5}) to prove Lemma \ref{L-g-state}. To do this we 
first establish the following auxiliary results (Lemmas \ref{L-concave-p}-\ref{L-minimum}). The proof of these lemmas is presented at the end of Appendix 3\footnote[2]{Lemma \ref{L-concave-p} is needed so as to prove Lemma \ref{L-feasible}. Lemmas \ref{L-concave-rho} and \ref{L-feasible} are needed so as to prove Lemma \ref{L-minimum}. Lemma \ref{L-minimum} is essential in establishing Lemma \ref{L-g-state}.}.

\begin{lemma}
	\label{L-concave-p}
	\normalfont
For any fixed time $t$ and any fixed message profile $m_{1:t}$, $W_t(\pi_t,m_{1:t},\allowbreak \boldsymbol{\rho})$ is a concave function of $\pi_t$. 
\end{lemma}

\begin{lemma}
	\label{L-concave-rho}
	\normalfont
For any fixed time $t$, any fixed message profile $m_{1:t}$, and any belief $\pi_t$, $W_t(\pi_t,m_{1:t},\boldsymbol{\rho})$ is a concave function of $\rho_{t+1}^{g,m_{1:t}}$. 
\end{lemma}

\begin{lemma}
	\label{L-feasible}
	\normalfont
In an obedient mechanism, we must have $\rho_{t+1}^{g,m_{1:t}} \in [\rho_{t+1}^{b,m_{1:t}},1]$, for any time $t$ and any message profile $m_{1:t}$.
\end{lemma}

\begin{lemma}
	\label{L-minimum}
	\normalfont
For any fixed time $t$, any fixed message profile $m_{1:t}$, and any belief $\pi_t$, we have
\begin{align}\label{eq:value-function14}
\argmin_{\rho_{t+1}^{g,m_{1:t}}}{W_t(\pi_t,m_{1:t},\boldsymbol{\rho})}=1.
\end{align}
\end{lemma}

Lemma \ref{L-minimum} shows that setting $\rho_{t+1}^{g,m_{1:t}}=1$ decreases the value function  $W_t(\pi_t,m_{1:t},\boldsymbol{\rho})$. Using this result, we complete the proof of Lemma \ref{L-g-state}, that is, we show that when the Markov chain is in the good state, irrespective of the message history, it is always optimal for the principal to recommend the detector to keep silent. That is 
\begin{align}\label{eq:rho-good2}
\rho_{t+1}^{* \hspace{0.05cm} g,m_{1:t}}=1, \forall t, m_{1:t}.
\end{align}

We prove this by contradiction. Consider an optimal recommendation policy $\boldsymbol{\rho}^*$ and assume that there exists at least one time $t$ and one message profile $m_{1:t}$ such that $\rho_{t+1}^{* \hspace{0.05cm} g,m_{1:t}} <1$. We construct another recommendation policy $\boldsymbol{\rho}$ such that $\rho_{t+1}^{b,m_{1:t}}=\rho_{t+1}^{* \hspace{0.05cm} b,m_{1:t}}$, $\rho_{t+1}^{g,m_{1:t}}=1$, and for all other times the policies $\boldsymbol{\rho}$ and $\boldsymbol{\rho}^*$ are the same. We show that the policy $\boldsymbol{\rho}$ leads to a higher expected utility for the principal and satisfies the obedience constraints corresponding
to situations where the detector is recommended to keep silent. Repeating the above argument for all times $s$ such that $\rho_{s+1}^{* \hspace{0.05cm} g,m_{1:s}} <1$ we construct a policy $\hat{\boldsymbol{\rho}}$ that satisfies $\hat{\rho}_{t+1}^{* \hspace{0.05cm} g,m_{1:t}}=1, \forall t \in \mathcal{T}$, it improves the principal's expected utility as compared to $\boldsymbol{\rho}^*$ and incentivizes the detector to keep silent when she is recommended to do so. In the new mechanism $\hat{\boldsymbol{\rho}}$, when the detector is recommended to declare the jump, she is sure that the Markov chain is in the bad state; hence, she will obey the recommendations for sure. Therefore, the mechanism  $\hat{\boldsymbol{\rho}}$ is a policy with higher expected utility than $\boldsymbol{\rho}^*$ that satisfies all the obedience constraints. This contradicts the optimality of $\boldsymbol{\rho}^*$.

To prove the claim stated above, we investigate the effect of setting $\rho_{t+1}^{g,m_{1:t}}$ to one on the detector's incentives to keep silent at times $t' \leq t$ and $t'>t$. 

\textbf{Times $t'$ where $t' \leq t$}: In this case, we show that setting $\rho_{t+1}^{g,m_{1:t}}=1$ reduces the value function at time $t'$. Therefore, if in the original mechanism we had $W_{t'}(\pi_{t'},m_{1:{t'}},\boldsymbol{\rho}^*)<\pi_{t'}$ (meaning that the detector had incentives to keep silent), in the new mechanism with $\rho_{t+1}^{g,m_{1:t}}=1$ this situation still holds. Therefore, the recommendation to keep silent at time $t'$ will be obeyed. We prove this claim by backward induction on time $t'$ as follows. 

\textbf{Basis of induction. $\boldsymbol{t'=t}$.} The result follows from Lemma \ref{L-minimum}.

\textbf{Induction step.} Suppose that setting $\rho_{t+1}^{g,m_{1:t}}=1$ decreases the value function $W_{k+1}(.)$, $k+1<t$. We prove that this is also true for time $k$. Consider the RHS of (\ref{eq:value-function5}) for time $k$ as follows:
\begin{align}\label{eq:value-function22}
\begin{split}
&W_k(\pi_k,m_{1:k},\boldsymbol{\rho})\hspace{-0.06cm}=\hspace{-0.06cm}\min{\left[\pi_k,c \hspace{0.05cm} (1\hspace{-0.06cm}-\hspace{-0.06cm}\pi_k)+(\pi_k(1\hspace{-0.06cm}-\hspace{-0.06cm}q) \rho_{k+1}^{g,m_{1:k}}\hspace{-0.06cm}+\hspace{-0.06cm}(1\hspace{-0.06cm}-\hspace{-0.06cm}\pi_k(1\hspace{-0.06cm}-\hspace{-0.06cm}q)) \rho_{k+1}^{b,m_{1:k}}) \right.} \\
&{\left. W_{k+1}(\frac{\pi_k(1-q) \rho_{k+1}^{g,m_{1:k}}}{\pi_k(1-q) \rho_{k+1}^{g,m_{1:k}}+(1-\pi_k(1-q)) \rho_{k+1}^{b,m_{1:k}}},m_{1:k},k,\boldsymbol{\rho})+ \right.} \\
&{\left. (\pi_k(1\hspace{-0.1cm}-\hspace{-0.1cm}q) (1\hspace{-0.1cm}-\hspace{-0.1cm}\rho_{k+1}^{g,m_{1:k}})\hspace{-0.1cm}+\hspace{-0.1cm}(1\hspace{-0.1cm}-\hspace{-0.1cm}\pi_k(1\hspace{-0.1cm}-\hspace{-0.1cm}q)) (1\hspace{-0.1cm}-\hspace{-0.1cm}\rho_{k+1}^{b,m_{1:k}})) \right.} \\
&{\left. W_{k+1}(\frac{\pi_k(1-q) (1-\rho_{k+1}^{g,m_{1:k}})}{\pi_k(1\hspace{-0.1cm}-\hspace{-0.1cm}q) (1\hspace{-0.1cm}-\hspace{-0.1cm}\rho_{k+1}^{g,m_{1:k}})\hspace{-0.1cm}+\hspace{-0.1cm}(1\hspace{-0.1cm}-\hspace{-0.1cm}\pi_k(1\hspace{-0.1cm}-\hspace{-0.1cm}q)) (1\hspace{-0.1cm}-\hspace{-0.1cm}\rho_{k+1}^{b,m_{1:k}})},\hspace{-0.05cm}m_{1:k},\hspace{-0.05cm}d,\hspace{-0.05cm}\boldsymbol{\rho})\right]}.
\end{split}
\end{align}
The first term of the minimum does not depend on $\rho_{t+1}^{g,m_{1:t}}$. The only component of the second term of the minimum that depends on $\rho_{t+1}^{g,m_{1:t}}$ is the value function $W_{k+1}(.)$. By the induction hypothesis, setting $\rho_{t+1}^{g,m_{1:t}}$ to one decreases $W_{k+1}(.)$ for any input, therefore, it decreases $W_{k}(.)$ (by Eq. (\ref{eq:value-function5})).

\textbf{Conclusion.} By the principle of induction, the statement is true for all $t' \leq t$. 

\textbf{Times $t'$ where $t' > t$}: Changing the value of $\rho_{t+1}^{g,m_{1:t}}$ with $m_{1:t} \neq (k)^{t}$ has no effect on the obedience constraints (\ref{eq:obedience-constraint-k})-(\ref{eq:obedience-constraint-d}) at time $t'> t$. However, setting $\rho_{t+1}^{g,(k)^{t}}=1$ increases the detector's belief in the good state of the Markov chain at time $t'$ when she is recommended to keep silent. This is because we have
\begin{align}\label{eq:value-function23}
\pi_{t'}=\mathbb{P}(s_{t'}\hspace{-0.05cm}=\hspace{-0.05cm}g|m_{1:t'},\boldsymbol{\rho})=\frac{\mathbb{P} (s_{t'}=g) \prod_{t^{''}=1}^{t'}{\rho_{t^{''}}^{g,(k)^{t''-1}}}}{\mathbb{P} (s_{t'}\hspace{-0.05cm}=\hspace{-0.05cm}g) \prod_{t^{''}=1}^{t'}{\rho_{t^{''}}^{g,(k)^{t''\hspace{-0.05cm}-\hspace{-0.05cm}1}}} \hspace{-0.05cm}+\hspace{-0.05cm} \mathbb{P} (s_{t'}\hspace{-0.05cm}=\hspace{-0.05cm}b, m_{1:t'}\hspace{-0.05cm}=\hspace{-0.05cm}(k)^{t'})}.
\end{align}
It is easy to show that the dependence of the RHS of (\ref{eq:value-function23}) on the parameter $\rho_{t+1}^{g,(k)^{t}}$ is of the form  
\begin{align}\label{eq:value-function24}
\frac{a \rho_{t+1}^{g,(k)^{t}}}{b \rho_{t+1}^{g,(k)^{t}} + d},
\end{align}
where $a,b,d$ are positive real numbers. This function is increasing in terms of $\rho_{t+1}^{g,(k)^{t}}$. Therefore, setting $\rho_{t+1}^{g,(k)^{t}}=1$ increases the belief of the detector in the good state of the Markov chain when she is recommended to keep silent.

We proved that for any (fixed) recommendation policy $\boldsymbol{\rho}$, at each time $t$, there is a threshold $l_{t}^{(k)^{t},\boldsymbol{\rho}}$ such that the detector keeps silent when she is recommended to do so if and only if $\mathbb{P}(s_t=g|m_{1:t}=(k)^{t},\boldsymbol{\rho}) > l_{t}^{(k)^{t},\boldsymbol{\rho}}$. In the optimal mechanism $\boldsymbol{\rho}^*$, the detector obeys the recommendation to keep silent at $t'$, i.e. $\mathbb{P}(s_{t'}=g|(k)^{t'},\boldsymbol{\rho}^*) > l_{t'}^{(k)^{t'},\boldsymbol{\rho}^*}$. In the modified mechanism $\boldsymbol{\rho}$, where $\rho_{t+1}^{g,(k)^{t}}=1$, we have $l_{t'}^{(k)^{t'},\boldsymbol{\rho}}=l_{t'}^{(k)^{t'},\boldsymbol{\rho}^*}$ for all $t' > t$ because for $t' > t$ the value function $W_{t'}(\pi_{t'},m_{1:{t'}},\boldsymbol{\rho})$ depends only on $\rho_{t'+1:T}$ and $\rho_{t'+1:T}=\rho^*_{t'+1:T}$. Since, by the arguments above, $\mathbb{P}(s_{t'}=g|(k)^{t'},\boldsymbol{\rho})> \mathbb{P}(s_{t'}=g|(k)^{t'},\boldsymbol{\rho}^*) > l_{t'}^{(k)^{t'},\boldsymbol{\rho}^*}=l_{t'}^{(k)^{t'},\boldsymbol{\rho}}$, under $\boldsymbol{\rho}$ the detector obeys the recommendation to remain silent at time $t'$. From the arguments above it follows that under $\boldsymbol{\rho}$ the detector always obeys the recommendation to keep silent.

Based on Eq. (\ref{eq:principal-utility}), we conclude that the principal's utility is an increasing function of the recommendation probabilities $\rho_{t+1}^{g,m_{1:t}}$. Therefore, the new constructed mechanism $\boldsymbol{\rho}$ results in a higher utility for the principal and the detector always obeys the recommendation to keep silent. Repeating the above argument for all times $s$ such that $\rho_{s+1}^{* \hspace{0.05cm} g,m_{1:s}} <1$ we construct a policy $\hat{\boldsymbol{\rho}}$ that satisfies $\hat{\rho}_{t+1}^{* \hspace{0.05cm} g,m_{1:t}}=1, \forall t \in \mathcal{T}$, it improves the principal's expected utility as compared to $\boldsymbol{\rho}^*$ and incentivizes the detector to keep silent when she is recommended to do so. In the new mechanism $\hat{\boldsymbol{\rho}}$, when the detector is recommended to declare the jump, she is sure that the Markov chain is in the bad state; hence, she will obey the recommendations for sure. Therefore, the mechanism  $\hat{\boldsymbol{\rho}}$ is a policy with higher expected utility than $\boldsymbol{\rho}^*$ that satisfies all the obedience constraints. This contradicts the optimality of $\boldsymbol{\rho}^*$ and proves Lemma \ref{L-g-state}.

\section{Proof of Lemma \ref{L-concave-p}}
We prove this by backward induction on time $t$. 

\textbf{Basis of induction.} Let $t=T$. According to (\ref{eq:value-function5}), for each message profile $m_{1:T}$, $W_T(\pi_T,m_{1:T},\boldsymbol{\rho})$ is the minimum of two affine, hence concave, functions of $\pi_T$. Minimum conserves concavity. Therefore, $W_T(\pi_T,m_{1:T},\boldsymbol{\rho})$ is a concave function of $\pi_T$.

\textbf{Induction step.} Suppose that $W_{t+1}(\pi_{t+1},m_{1:t+1},\boldsymbol{\rho})$ is a concave function of $\pi_{t+1}$. Now, we want to show this is also true for $t$. Concave functions can be written as the infimum of affine functions; so we have
\begin{align}\label{eq:value-function6}
W_{t+1}(\pi_{t+1},m_{1:t+1},\boldsymbol{\rho})=\inf_{i}{\left[\alpha_i^{m_{1:t+1}}\pi_{t+1}+\beta_i^{m_{1:t+1}}\right]}.
\end{align}
Substituting (\ref{eq:value-function6}) in (\ref{eq:value-function5}), we have
\begin{align}\label{eq:value-function7}
\begin{split}
&W_t(\pi_t,m_{1:t},\boldsymbol{\rho})=\min{\left[\pi_t,c \hspace{0.05cm} (1-\pi_t)+ \inf_{i}{\left\{\alpha_i^{m_{1:t},k}\pi_t(1-q) \rho_{t+1}^{g,m_{1:t}}+ \right.}\right.} \\
&{\left. {\left. \beta_i^{m_{1:t},k}(\pi_t(1-q) \rho_{t+1}^{g,m_{1:t}}+(1-\pi_t(1-q)) \rho_{t+1}^{b,m_{1:t}})\right\}} +  \inf_{i}{\left\{\alpha_i^{m_{1:t},d}\pi_t(1-q) \right.}\right.} \\
&{\left. {\left. (1\hspace{-0.05cm}-\hspace{-0.05cm}\rho_{t+1}^{g,m_{1:t}}) \hspace{-0.05cm}+\hspace{-0.05cm}\beta_i^{m_{1:t},d}(\pi_t(1\hspace{-0.05cm}-\hspace{-0.05cm}q) (1-\rho_{t+1}^{g,m_{1:t}})+(1-\pi_t(1-q)) (1-\rho_{t+1}^{b,m_{1:t}}))\right\}}\right]}.
\end{split}
\end{align}
It can be seen that the terms inside of both infima are affine in $\pi_t$, hence concave. Therefore, since the infimum preserve concavity, each term inside the minimum is concave in $\pi_t$. Using the concavity preserving of the minimum function proves the concavity of the value function $W_t(\pi_t,m_{1:t},\boldsymbol{\rho})$ in terms of $\pi_t$.

\textbf{Conclusion.} By the principle of induction, the statement is true for all $t=1, \ldots, T$. This completes the proof of lemma \ref{L-concave-p}.

\section{Proof of Lemma \ref{L-concave-rho}}
The assertion of this lemma easily follows from (\ref{eq:value-function7}).

\section{Proof of Lemma \ref{L-feasible}}
In Lemma \ref{L-concave-p} we proved that: at any time $t$, the expected cost of the detector due to her decision to stop and declare the jump ($a_t=1$) is given by the first term of the RHS of (\ref{eq:value-function7}) and  is a linear increasing function of $\pi_t$; the expected cost due to her decision to remain silent is given by the second term of the RHS of (\ref{eq:value-function7}) and is a concave function of $\pi_t$. Denote the above expected costs by $V_t(\pi_t,m_{1:t},\boldsymbol{\rho},a_t=1)$ and $V_t(\pi_t,m_{1:t},\boldsymbol{\rho},a_t=0)$, respectively.

Now, we claim that if 
\begin{align}\label{eq:value-function8}
V_t(\pi,m_{1:t},\boldsymbol{\rho},a_t=0)<V_t(\pi,m_{1:t},\boldsymbol{\rho},a_t=1)
\end{align}
for some $\pi$, then (\ref{eq:value-function8}) is also true for all $\pi'> \pi$.

We prove the claim by contradiction. Suppose that 
\begin{align}\label{eq:value-function9}
V_t(\pi,m_{1:t},\boldsymbol{\rho},a_t=0)<V_t(\pi,m_{1:t},\boldsymbol{\rho},a_t=1), \tag{*}
\end{align}
and there exists a belief $\pi'> \pi$ where 
\begin{align}\label{eq:value-function10}
V_t(\pi',m_{1:t},\boldsymbol{\rho},a_t=0)>V_t(\pi',m_{1:t},\boldsymbol{\rho},a_t=1). \tag{**}
\end{align}
We know that 
\begin{align}\label{eq:value-function11}
V_t(0,m_{1:t},\boldsymbol{\rho},a_t=0)=c>0=V_t(0,m_{1:t},\boldsymbol{\rho},a_t=1). \tag{***}
\end{align}
Therefore, (\ref{eq:value-function9})-(\ref{eq:value-function11}) show that the graphs of functions $V_t(\pi_t,m_{1:t},\boldsymbol{\rho},a_t=0)$ and $V_t(\pi_t,m_{1:t},\boldsymbol{\rho},a_t=1)$ have two intersections $l^1 \in (0,\pi)$ and $l^2 \in (\pi,\pi')$, where the graph of $V_t(\pi_t,m_{1:t},\boldsymbol{\rho},a_t=0)$ is below the line $V_t(\pi_t,m_{1:t},\boldsymbol{\rho},a_t=1)$ in the interval $(l^1,l^2)$. This contradicts the concavity of the function $V_t(\pi_t,\allowbreak m_{1:t},\allowbreak \boldsymbol{\rho},a_t=0)$ and hence the claim is true. The truth of this claim shows that at any time $t$ and for any fixed $m_{1:t}$ and $\boldsymbol{\rho}$, the detector's optimal policy is of threshold type with respect to $\pi_t$; that is there is a threshold $l_{t}^{m_{1:t},\boldsymbol{\rho}}$ such that the detector declares the jump if and only if her belief $\pi_t$ about the good state of the Markov chain is below the threshold $l_{t}^{m_{1:t},\boldsymbol{\rho}}$. Therefore, for a mechanism to be obedient we need to have the following property:
\begin{align}\label{eq:value-function12}
\mathbb{P} (s_{t+1}=g |m_{1:t},\boldsymbol{\rho}, m_{t+1}=d) \leq \mathbb{P} (s_{t+1}=g |m_{1:t},\boldsymbol{\rho}, m_{t+1}=k), \forall t, m_{1:t}.
\end{align}
Writing the probabilities appearing in (\ref{eq:value-function12}) in terms of the belief of time $t$, we can simplify the necessary condition (\ref{eq:value-function12}) as follows:
\begin{align}\label{eq:value-function13}
&\frac{\pi_t (1-q) (1-\rho_{t+1}^{g,m_{1:t}})}{\pi_t (1-q) (1-\rho_{t+1}^{g,m_{1:t}})+(1-\pi_t (1-q)) (1-\rho_{t+1}^{b,m_{1:t}})} \leq \nonumber\\
&\frac{\pi_t (1-q) \rho_{t+1}^{g,m_{1:t}}}{\pi_t (1-q) \rho_{t+1}^{g,m_{1:t}}+(1-\pi_t (1-q)) \rho_{t+1}^{b,m_{1:t}}}, \quad \forall t, m_{1:t}.
\end{align}
Simplifying the above expression gives us $\rho_{t+1}^{g,m_{1:t}} \geq \rho_{t+1}^{b,m_{1:t}}$ as a necessary condition for an obedient mechanism, and hence completes the proof of Lemma \ref{L-feasible}.

\section{Proof of Lemma \ref{L-minimum}}
According to Lemma \ref{L-concave-rho}, $W_t(\pi_t,m_{1:t},\boldsymbol{\rho})$ is a concave function of $\rho_{t+1}^{g,m_{1:t}}$. Therefore, the minimum of the function is attained either at the beginning or at the end of the feasible interval of the parameter $\rho_{t+1}^{g,m_{1:t}}$ which is derived in Lemma \ref{L-feasible} as $[\rho_{t+1}^{b,m_{1:t}},1]$. Therefore, to prove Lemma \ref{L-minimum}, we only need to show that 
\begin{align}\label{eq:value-function15}
W_t(\pi_t,m_{1:t},\hspace{-0.03cm}\boldsymbol{\rho}\backslash\{\rho_{t+1}^{g,m_{1:t}}\},\rho_{t+1}^{g,m_{1:t}}\hspace{-0.11cm}=\hspace{-0.09cm}1)\hspace{-0.05cm} \leq \hspace{-0.05cm} W_t(\pi_t,\hspace{-0.03cm}m_{1:t},\boldsymbol{\rho}\backslash\{\rho_{t+1}^{g,m_{1:t}}\},\rho_{t+1}^{g,m_{1:t}}\hspace{-0.11cm}=\hspace{-0.09cm}\rho_{t+1}^{b,m_{1:t}}),
\end{align}
for every time $t$, belief $\pi_t$, and message profile $m_{1:t}$, where $\boldsymbol{\rho}\backslash\{\rho_{t+1}^{g,m_{1:t}}\}$ denotes the recommendation policy $\boldsymbol{\rho}$ excluding the element $\rho_{t+1}^{g,m_{1:t}}$. Using (\ref{eq:value-function5}), we have
\begin{align}\label{eq:value-function16}
\begin{split}
&W_t(\pi_t,m_{1:t},\boldsymbol{\rho}\backslash \{\rho_{t+1}^{g,m_{1:t}}\},\rho_{t+1}^{g,m_{1:t}}=1)= \\
&\min{\left[\pi_t,c \hspace{0.05cm} (1-\pi_t)+ (\pi_t(1\hspace{-0.1cm}-\hspace{-0.1cm}q)+(1\hspace{-0.1cm}-\hspace{-0.1cm}\pi_t(1\hspace{-0.1cm}-\hspace{-0.1cm}q)) \rho_{t+1}^{b,m_{1:t}}) \right.} \\
&{\left. W_{t+1}(\frac{\pi_t(1-q)}{\pi_t(1\hspace{-0.1cm}-\hspace{-0.1cm}q)+(1\hspace{-0.1cm}-\hspace{-0.1cm}\pi_t(1\hspace{-0.1cm}-\hspace{-0.1cm}q)) \rho_{t+1}^{b,m_{1:t}}},m_{1:t},k,\boldsymbol{\rho}\backslash\{\rho_{t+1}^{g,m_{1:t}}\},\rho_{t+1}^{g,m_{1:t}}=1) + \right.} \\
&{\left. (1-\pi_t(1-q)) (1-\rho_{t+1}^{b,m_{1:t}}) W_{t+1}(0,m_{1:t},d,\boldsymbol{\rho}\backslash\{\rho_{t+1}^{g,m_{1:t}}\},\rho_{t+1}^{g,m_{1:t}}=1)\right]}.
\end{split}
\end{align}
When the detector is sure that the state is bad, she will declare the jump irrespective of the message she has received, and incurs no cost. Therefore, we have
\begin{align}\label{eq:value-function16-1}
&W_{t+1}(0,m_{1:t},d,\boldsymbol{\rho}\backslash\{\rho_{t+1}^{g,m_{1:t}}\},\rho_{t+1}^{g,m_{1:t}}=1)=\nonumber\\
&W_{t+1}(0,m_{1:t},k,\boldsymbol{\rho}\backslash\{\rho_{t+1}^{g,m_{1:t}}\},\rho_{t+1}^{g,m_{1:t}}=1)=0.
\end{align}
This equality allows us to replace $W_{t+1}(0,m_{1:t},d,\boldsymbol{\rho}\backslash\{\rho_{t+1}^{g,m_{1:t}}\},\rho_{t+1}^{g,m_{1:t}}=1)$ in (\ref{eq:value-function16}) by $W_{t+1}(0,m_{1:t},k,\boldsymbol{\rho}\backslash\{\rho_{t+1}^{g,m_{1:t}}\},\allowbreak \rho_{t+1}^{g,m_{1:t}}=1)$ and get 
\begin{align}\label{eq:value-function17}
\begin{split}
&W_t(\pi_t,m_{1:t},\boldsymbol{\rho}\backslash \{\rho_{t+1}^{g,m_{1:t}}\},\rho_{t+1}^{g,m_{1:t}}=1)= \\
&\min{\left[\pi_t,c \hspace{0.05cm} (1-\pi_t)+ (\pi_t(1\hspace{-0.1cm}-\hspace{-0.1cm}q)+(1\hspace{-0.1cm}-\hspace{-0.1cm}\pi_t(1\hspace{-0.1cm}-\hspace{-0.1cm}q)) \rho_{t+1}^{b,m_{1:t}}) \right.} \\
&{\left. W_{t+1}(\frac{\pi_t(1-q)}{\pi_t(1\hspace{-0.1cm}-\hspace{-0.1cm}q)+(1\hspace{-0.1cm}-\hspace{-0.1cm}\pi_t(1\hspace{-0.1cm}-\hspace{-0.1cm}q)) \rho_{t+1}^{b,m_{1:t}}},m_{1:t},k,\boldsymbol{\rho}\backslash\{\rho_{t+1}^{g,m_{1:t}}\},\rho_{t+1}^{g,m_{1:t}}=1) + \right.} \\
&{\left. (1-\pi_t(1-q)) (1-\rho_{t+1}^{b,m_{1:t}}) W_{t+1}(0,m_{1:t},k,\boldsymbol{\rho}\backslash\{\rho_{t+1}^{g,m_{1:t}}\},\rho_{t+1}^{g,m_{1:t}}=1)\right]}.
\end{split}
\end{align}
It can be concluded from (\ref{eq:value-function-T}) and (\ref{eq:value-function5}) that for any $t$ and any fixed $\pi_{t}$ and $m_{1:t}$, the value function $W_{t}(\pi_{t},m_{1:t},\boldsymbol{\rho})$ is independent of the recommendation policies $\rho_{t'}$, $t' \leq t$ (the effect of $\rho_{t'}$, $t' \leq t$, on $W_{t}(\pi_{t},m_{1:t},\boldsymbol{\rho})$ is captured/summarized by $\pi_{t}$ and $m_{1:t}$.). Therefore, we can write (\ref{eq:value-function17}) as
\begin{align}\label{eq:value-function17-1}
\begin{split}
&W_t(\pi_t,m_{1:t},\boldsymbol{\rho}\backslash \{\rho_{t+1}^{g,m_{1:t}}\},\rho_{t+1}^{g,m_{1:t}}=1)=\\
&\min{\left[\pi_t,c \hspace{0.05cm} (1-\pi_t)+ (\pi_t(1-q)+(1-\pi_t(1-q)) \rho_{t+1}^{b,m_{1:t}})\right.} \\
&{\left. W_{t+1}(\frac{\pi_t(1-q)}{\pi_t(1-q)+(1-\pi_t(1-q)) \rho_{t+1}^{b,m_{1:t}}},m_{1:t},k,\boldsymbol{\rho}) + \right.} \\
&{\left. (1-\pi_t(1-q)) (1-\rho_{t+1}^{b,m_{1:t}}) W_{t+1}(0,m_{1:t},k,\boldsymbol{\rho})\right]}.
\end{split}
\end{align}
When $\rho_{t+1}^{g,m_{1:t}}=\rho_{t+1}^{b,m_{1:t}}$, we have from (\ref{eq:value-function5}) that
\begin{align}\label{eq:value-function18}
\begin{split}
&W_t(\pi_t,m_{1:t},\boldsymbol{\rho}\backslash \{\rho_{t+1}^{g,m_{1:t}}\},\rho_{t+1}^{g,m_{1:t}}=
\rho_{t+1}^{b,m_{1:t}})=\min{\left[\pi_t,c \hspace{0.05cm} (1-\pi_t)+\right.} \\
&{\left. \rho_{t+1}^{b,m_{1:t}} W_{t+1}(\pi_t(1-q),m_{1:t},k,\boldsymbol{\rho}) + (1-\rho_{t+1}^{b,m_{1:t}})  W_{t+1}(\pi_t(1-q),m_{1:t},d,\boldsymbol{\rho})\right]}.
\end{split}
\end{align}
In this case, the message sent by the principal at time $t$ is independent of the observed state. Therefore, the detector neglects this data in her future decisions. Thus, we have 
\begin{align}\label{eq:value-function19}
W_{t+1}(\pi_t(1-q),m_{1:t},k,\boldsymbol{\rho})=W_{t+1}(\pi_t(1-q),m_{1:t},d,\boldsymbol{\rho}).
\end{align}
Substituting (\ref{eq:value-function19}) into (\ref{eq:value-function18}) gives
\begin{align}\label{eq:value-function20}
&W_t(\pi_t,m_{1:t},\boldsymbol{\rho}\backslash \{\rho_{t+1}^{g,m_{1:t}}\},\rho_{t+1}^{g,m_{1:t}}=\rho_{t+1}^{b,m_{1:t}})=\nonumber\\
&\min{\left[\pi_t,c \hspace{0.05cm} (1-\pi_t)+W_{t+1}(\pi_t(1-q),m_{1:t},k,\boldsymbol{\rho})\right]}.
\end{align}
According to Lemma \ref{L-concave-p}, $W_{t+1}(\pi_t(1-q),m_{1:t},k,\boldsymbol{\rho})$ is a concave function of its first element. Thus, we have
\begin{align}\label{eq:value-function21}
&(\pi_t(1\hspace{-0.09cm}-\hspace{-0.07cm}q)\hspace{-0.09cm}+\hspace{-0.09cm}(1\hspace{-0.09cm}-\hspace{-0.07cm}\pi_t(1\hspace{-0.09cm}-\hspace{-0.07cm}q)) \rho_{t+1}^{b,m_{1:t}}) W_{t\hspace{-0.03cm}+\hspace{-0.03cm}1}(\frac{\pi_t(1-q)}{\pi_t(1\hspace{-0.09cm}-\hspace{-0.07cm}q)+(1\hspace{-0.09cm}-\hspace{-0.07cm}\pi_t(1\hspace{-0.07cm}-\hspace{-0.07cm}q)) \rho_{t+1}^{b,m_{1:t}}},m_{1:t},k,\boldsymbol{\rho}) \nonumber\\
&+(1-\pi_t(1-q)) (1-\rho_{t+1}^{b,m_{1:t}}) W_{t+1}(0,m_{1:t},k,\boldsymbol{\rho}) \leq 
W_{t+1}(\pi_t(1-q),m_{1:t},k,\boldsymbol{\rho}).
\end{align}
Comparing (\ref{eq:value-function17-1}) and (\ref{eq:value-function20}) based on the result derived in (\ref{eq:value-function21}) proves that the inequality (\ref{eq:value-function15}) holds; hence the statement of Lemma \ref{L-minimum} is true.

\section{Proof of Lemma \ref{L-priority}}
We show that for each optimal sequential information disclosure mechanism $\boldsymbol{\rho}$ there is a time-based prioritized mechanism which obtains the same expected utility for the principal and satisfies all the obedience constraints.

Based on Lemma \ref{L-g-state}, we set $\rho_{t}^{g,(k)^{t-1}}=1$, for all $t$, $t=1,2,\ldots, T$. Furthermore, we let $\rho_{t}^{s_t,m_{1:t-1}}$ be arbitrary $0 \leq \rho_{t}^{s_t,m_{1:t-1}} \leq 1$, whenever $m_{1:t-1} \neq (k)^{t-1}$. Thus, we concentrate on $\rho_{t}^{b,(k)^{t-1}}$, $t=1,2,\ldots, T$. For ease of notation, throughout the remainder of the proof we denote $\rho_{t}^{b,(k)^{t-1}}$ by $\rho_t$.

Consider an optimal mechanism $\boldsymbol{\rho}$ and let time $t$ be the last time period where $\rho_t <1$ and $\rho_{t+1} >0$. Using the result of Lemma \ref{L-g-state} and the new notation, we find that the principal's expected utility according to $\boldsymbol{\rho}$ is
\begin{align}\label{eq:proof-priority1}
\mathbb{E}^{\boldsymbol{\rho}} \{U^P\}=
\sum_{l=1}^T{\sum_{\theta'=1}^{l}{\mathbb{P}(\theta=\theta') \prod_{t'=\theta'}^{l}{\rho_{t'}}}}+\sum_{l=1}^T{\sum_{\theta'=l+1}^{T+1}{\mathbb{P}(\theta=\theta')}}.
\end{align}
It is clear from (\ref{eq:proof-priority1}) that the principal's utility is a continuous and increasing function of each $\rho_{t'}$, $t'=1, 2, \ldots, T$. Since $\rho_{t+1} >0$ there exists a $\gamma>0$ such that $\hat{\rho}_{t+1}=\rho_{t+1}-\gamma \geq 0$. Replacing $\rho_{t+1}$ by $\hat{\rho}_{t+1}$ reduces the expected utility of the principal. However, since $\mathbb{E} \{U^P(\tau)\}$ is a continuously increasing function of $\rho_{t}$, there exists some $\epsilon(\gamma)>0$ such that increasing $\rho_{t}$ by the amount of $\epsilon(\gamma)$ can compensate this decrease. This compensation is feasible if $\hat{\rho}_{t}=\rho_{t}+\epsilon(\gamma)$ does not exceed its upper limit $1$. To ensure this, we take $\gamma=\min{(\epsilon^{-1}(1-\rho_t),\rho_{t+1})}$, where $\epsilon^{-1}(.)$ is the inverse of function $\epsilon(.)$. The function $\epsilon(.)$ is one to one and increasing in $\gamma$, hence it has an inverse \footnote[2]{The explicit declaration of function $\epsilon(.)$ can be derived from equation (\ref{eq:proof-priority6}) that will be derived later.}. Choosing this $\gamma$ results in either $\hat{\rho}_t =1$ or  $\hat{\rho}_{t+1}=0$.

The arguments above show that the principal's expected utility when he discloses his information based on the mechanism $\boldsymbol{\hat{\rho}}$, where $\hat{\rho}_{t'}=\rho_{t'}$, for all $t' \neq t, t+1$, and
\begin{align}\label{eq:proof-priority2}
\hat{\rho}_{t}=\rho_{t}+\epsilon(\gamma), \hspace{0.3cm} \hat{\rho}_{t+1}=\rho_{t+1}-\gamma,
\end{align}
is the same as the average utility he gets when he uses the optimal mechanism $\boldsymbol{\rho}$; i.e.
\begin{align}\label{eq:proof-priority3}
\sum_{l=1}^T{\sum_{\theta'=1}^{l}{\mathbb{P}(\theta=\theta') \prod_{t'=\theta'}^l{\hat{\rho}_{t'}}}}=\sum_{l=1}^T{\sum_{\theta'=1}^{l}{\mathbb{P}(\theta=\theta') \prod_{t'=\theta'}^l{\rho_{t'}}}}.
\end{align}
Therefore, showing the new mechanism $\boldsymbol{\hat{\rho}}$ satisfies the obedience constraints will prove its optimality. 

Since the recommendation to declare that the jump has occurred is always obeyed, we only need to investigate the obedience constraints (\ref{eq:obedience-constraint-k}) when the detector is recommended to keep silent. Using the results derived in (\ref{eq:obedience-constraint-k-2}) and (\ref{eq:future_cost-obey2}), we can show that the mechanism $\boldsymbol{\hat{\rho}}$ satisfies the obedience constraints if and only if 
\begin{align}\label{eq:proof-priority4}
c \sum_{l=t''}^T{\sum_{\theta'=1}^{l}{\mathbb{P} (\theta=\theta')\prod_{t'=\theta'}^{l}{\hat{\rho}_{t'}}}} \leq
1-\sum_{\theta'=1}^{t''}{\mathbb{P} (\theta=\theta')}, \forall t'' \in \mathcal{T}.
\end{align}
Using (\ref{eq:proof-priority3}) we can show that for $t'' \leq t$, we have
\begin{align}\label{eq:proof-priority5}
c \sum_{l=t''}^T{\sum_{\theta'=1}^{l}{\mathbb{P} (\theta=\theta')\prod_{t'=\theta'}^{l}{\hat{\rho}_{t'}}}}= c \sum_{l=t''}^T{\sum_{\theta'=1}^{l}{\mathbb{P} (\theta=\theta')\prod_{t'=\theta'}^{l}{\rho_{t'}}}} \leq
1-\sum_{\theta'=1}^{t''}{\mathbb{P} (\theta=\theta')},
\end{align}
where the last inequality follows from the obedience property of the original mechanism $\boldsymbol{\rho}$. Therefore, the obedience constraint for all times before $t$ is satisfied. To show this is also true for times after $t$ we need to take a closer look at (\ref{eq:proof-priority3}). Canceling out the equal terms from both sides of the equality derived in (\ref{eq:proof-priority3}), we get
\begin{align}\label{eq:proof-priority6}
&\epsilon(\gamma)\sum_{\theta'=1}^{t}{\mathbb{P}(\theta=\theta') \prod_{t'=\theta'}^{t-1}{\rho_{t'}}}+\Big(-\gamma\mathbb{P}(\theta=t+1)+\nonumber\\
&(\epsilon(\gamma)\rho_{t+1}-\gamma \rho_t -\gamma \epsilon(\gamma))\sum_{\theta'=1}^{t}{\mathbb{P}(\theta=\theta') \prod_{t'=\theta'}^{t-1}{\rho_{t'}}}\Big)\sum_{l=t+1}^T{ \prod_{t'=t+2}^l{\rho_{t'}}}=0.
\end{align}
The first term in (\ref{eq:proof-priority6}) and $\sum_{l=t+1}^T{ \prod_{t'=t+2}^l{\rho_{t'}}}$ are both non-negative. Therefore, we have
\begin{align}\label{eq:proof-priority7}
-\gamma\mathbb{P}(\theta=t+1)+
(\epsilon(\gamma)\rho_{t+1}-\gamma \rho_t -\gamma \epsilon(\gamma))\sum_{\theta'=1}^{t}{\mathbb{P}(\theta=\theta') \prod_{t'=\theta'}^{t-1}{\rho_{t'}}}\leq 0.
\end{align}
By substituting (\ref{eq:proof-priority2}) in the left hand side of the obedience constraint (\ref{eq:proof-priority4}) for time $t''\geq t+1$, we obtain
\begin{align}\label{eq:proof-priority8}
&c \sum_{l=t''}^T{\sum_{\theta'=1}^{l}{\mathbb{P} (\theta=\theta')\prod_{t'=\theta'}^{l}{\hat{\rho}_{t'}}}}=c \sum_{l=t''}^T{\sum_{\theta'=1}^{l}{\mathbb{P} (\theta=\theta')\prod_{t'=\theta'}^{l}{\rho_{t'}}}}+ \nonumber\\
&\Big(\hspace{-0.05cm}-\hspace{-0.05cm}\gamma\mathbb{P}(\theta\hspace{-0.05cm}=\hspace{-0.05cm}t\hspace{-0.05cm}+\hspace{-0.05cm}1)\hspace{-0.07cm}+\hspace{-0.07cm}(\epsilon(\gamma)\rho_{t+1}\hspace{-0.05cm}-\hspace{-0.05cm}\gamma \rho_t \hspace{-0.05cm}-\hspace{-0.05cm}\gamma \epsilon(\gamma))\sum_{\theta'=1}^{t}{\mathbb{P}(\theta\hspace{-0.05cm}=\hspace{-0.05cm}\theta')\hspace{-0.05cm} \prod_{t'=\theta'}^{t-1}{\rho_{t'}}}\Big) \sum_{l=t''}^T{\hspace{-0.08cm} \prod_{t'=t+2}^l{\rho_{t'}}} \hspace{-0.05cm} \leq \nonumber \\
& c \sum_{l=t''}^T{\sum_{\theta'=1}^{l}{\mathbb{P} (\theta=\theta')\prod_{t'=\theta'}^{l}{\rho_{t'}}}} \leq 1-\sum_{\theta'=1}^{t''}{\mathbb{P} (\theta=\theta')},
\end{align}
where the first inequality is based on (\ref{eq:proof-priority7}), and the second one follows from the fact that the original mechanism $\boldsymbol{\rho}$ is obedient, hence it satisfies the obedience constraints (\ref{eq:proof-priority4}). The results derived in (\ref{eq:proof-priority5}) and (\ref{eq:proof-priority8}) show that the new mechanism $\boldsymbol{\hat{\rho}}$ satisfies the obedience constraints, hence it is optimal. In this new mechanism, we have either $\hat{\rho}_t =1$ or  $\hat{\rho}_{t+1}=0$. Therefore, irrespective of the original mechanism $\boldsymbol{\rho}$, the new (modified) mechanism satisfies the condition of having a time-based priority for recommending the detector to keep silent at time $t$ and at the next time period $t+1$. 

It is clear from the above arguments that by repeating this procedure for any time $t$ that violates the priority condition in a backward direction we can construct a time-based prioritized mechanism which is optimal. This completes the proof of Lemma \ref{L-priority}.

\section{Proof of Theorem \ref{T-binding}}
After removing the terms in (\ref{eq:binding1}) that do not depend on $t$ and doing some simple algebra, we have
\begin{align}\label{eq:binding4}
\begin{split}
&\argmin_{t \leq n_p}{q_{n_p}^t}=\argmin_{t \leq n_p}{\left[\mathbb{P}(\theta > t)-c \sum_{l=t}^{n_p-1}{\mathbb{P}(\theta \leq l)}\right]}=\\
&\argmin_{t \leq n_p}{\left[\mathbb{P}(\theta > t)+c \sum_{l=1}^{n_p-1}{\mathbb{P}(\theta \leq l)}-c \sum_{l=t}^{n_p-1}{\mathbb{P}(\theta \leq l)}\right]}=\\
&\argmin_{t \leq n_p}{\left[\mathbb{P}(\theta > t)+c \sum_{l=1}^{t-1}{\mathbb{P}(\theta \leq l)}\right]}\\
&\argmin_{t \leq n_p}{\mathbb{E}^{No} \{J^D(t,\theta)\}}.
\end{split}
\end{align}
In Feature 2, we showed that $J^D(t,\theta)$ attains its minimum at $t=\tau^{No}$. Therefore, since $n_p \geq \tau^{No}$, we have 
\begin{align}\label{eq:binding5}
\argmin_{t \leq n_p}{q_{n_p}^t}=\tau^{No}.
\end{align}
This completes the proof of Theorem \ref{T-binding}.

\bibliographystyle{apalike} 
\bibliography{bib}

\end{document}